\documentclass[sigconf]{acmart}

\usepackage{soul}
\usepackage[ruled,vlined]{algorithm2e}

\usepackage{array}
\makeatletter
\newcolumntype{"}{@{\hskip\tabcolsep\vrule width 2pt\hskip\tabcolsep}}
\makeatother

\usepackage{graphicx}
\usepackage{balance}  % for  \balance command ON LAST PAGE  (only there!)

\usepackage{url}
\usepackage[shortlabels]{enumitem}
\usepackage{amsmath}
\usepackage{wasysym}
\usepackage{textcomp}
\usepackage[T1]{fontenc}
\usepackage[noend]{algpseudocode}
\usepackage[caption=false,font=footnotesize]{subfig}
\usepackage[belowskip=-8pt,aboveskip=5pt]{caption}
\usepackage[belowskip=0pt,aboveskip=4pt]{caption}
\usepackage[aboveskip=0pt]{caption}
\usepackage[belowskip=0pt]{caption}
\usepackage{hhline}
\usepackage{makecell}

\usepackage[compact]{titlesec}
\titlespacing{\section}{0pt}{1ex}{0ex}
\titlespacing{\subsection}{0pt}{1ex}{0ex}
\titlespacing{\subsubsection}{0pt}{0ex}{0ex}

\AtBeginDocument{%
  \providecommand\BibTeX{{%
    \normalfont B\kern-0.5em{\scshape i\kern-0.25em b}\kern-0.8em\TeX}}}

\begin{document}

\newpage
%\title{Why Do My Blockchain Transactions Fail? A Study of Hyperledger Fabric [Experiment and Analysis]}
%\title{BlockProM: An optimization recommendation strategy for blockchains using process mining}
%\title{BlockProM: An Optimization Recommender for Blockchains Based on Process Mining}
%\title{MulBORe: A Multi-Level Blockchain Optimization Recommender}
%\title{BlockOptR: A Multi-Level Blockchain Optimization Recommender}
\title{How To Optimize My Blockchain?  \protect\\ A Multi-Level Recommendation Approach}

\author{Jeeta Ann Chacko}
\email{chacko@in.tum.de}
\affiliation{%
  %\institution{Middleware Systems Research Group}
  \institution{Technical University of Munich}
}
\author{Ruben Mayer}
\email{mayerr@in.tum.de}
\affiliation{%
  %\institution{Middleware Systems Research Group}
  \institution{Technical University of Munich}
}
\author{Hans-Arno Jacobsen}
\email{jacobsen@eecg.toronto.edu}
\affiliation{%
  %\institution{Middleware Systems Research Group}
  \institution{University of Toronto}
}


\begin{abstract}
   Aside from the conception of new blockchain architectures, existing blockchain optimizations in the literature primarily focus on system or data-oriented optimizations within prevailing  blockchains. However, since blockchains handle multiple aspects ranging from organizational governance to smart contract design, a holistic approach that encompasses all the different layers of a given blockchain system is required to ensure that all optimization opportunities are taken into consideration. In this vein, we define a multi-level optimization recommendation approach that identifies optimization opportunities within a blockchain at the system, data, and user level. Multiple metrics and attributes are derived from a blockchain log and nine optimization recommendations are formalized. We implement an automated optimization recommendation tool, \textsf{BlockOptR}, based on these concepts. The system is extensively evaluated with a wide range of workloads covering multiple real-world scenarios. After implementing the recommended optimizations, we observe an average of 20\% improvement in the success rate of transactions and an average of 40\% improvement in latency.
\end{abstract}

%%\begin{CCSXML}
%<ccs2012>
%   <concept>
%       <concept_id>10002951.10002952</concept_id>
%%       <concept_desc>Information systems~Data management systems</concept_desc>
 %      <concept_significance>500</concept_significance>
 %      </concept>
 %</ccs2012>
%\end{CCSXML}

%\ccsdesc[500]{Information systems~Data management systems}

%\keywords{blockchains, transaction failures, concurrency}

\maketitle
\fancyhead{}
%\linenumbers

%\balance
%(B\lowercase{lock}P\lowercase{ro}M)

\section{Introduction}
\label{sec:introduction}

When blockchains were first introduced, they supported only simple cryptocurrency exchange transactions~\cite{nakamoto2008bitcoin}. However, over time blockchains evolved to support complex transactions using smart contracts, thus entering the arena of decentralized transactional management systems such as distributed databases~\cite{10.1145/3448016.3452789}. Since blockchains target consensus in a trustless environment, they cannot easily match the performance of databases~\cite{10.14778/3342263.3342632, Falazi2019TransactionalPropsPB, 8456055, 8525392, 9143893, 8246573, https://doi.org/10.1002/cpe.5227}. However, with the advent of permissioned blockchains that offer access control and transaction execution policies, block\-chains strive to improve their performance while still providing at least partially decentralized trust~\cite{Androulaki:2018:HFD:3190508.3190538, morgan2016quorum, greenspan2015multichain, 10.1145/3448016.3457539}.

Apart from the proliferation of new blockchain system designs, there is highly vibrant and diverse ongoing research in the domain of system optimizations that focus on performance enhancements within prevailing permissioned blockchains~\cite{Sharma:2019:BLB:3299869.3319883, 10.1145/3318464.3389693, gorenflo2019fastfabric, istvan2018streamchain, 8526892, 10.1145/3448016.3452823, 8843212, 9306742, 8636255, 9569819}. The vast range of the literature targets control parameter tuning~\cite{8526892, 10.1145/3448016.3452823, 8636255}, transaction execution remodeling~\cite{gorenflo2019fastfabric, 8843212, 9306742}, and smart contract optimization~\cite{9569819}. However, we notice that a collective approach that encompasses all these optimization possibilities for a particular blockchain under the same umbrella is missing. 
%system redesign~\cite{Sharma:2019:BLB:3299869.3319883, 10.1145/3318464.3389693, istvan2018streamchain},
\begin{figure}[t]

\centering
\includegraphics[width=0.65\columnwidth]{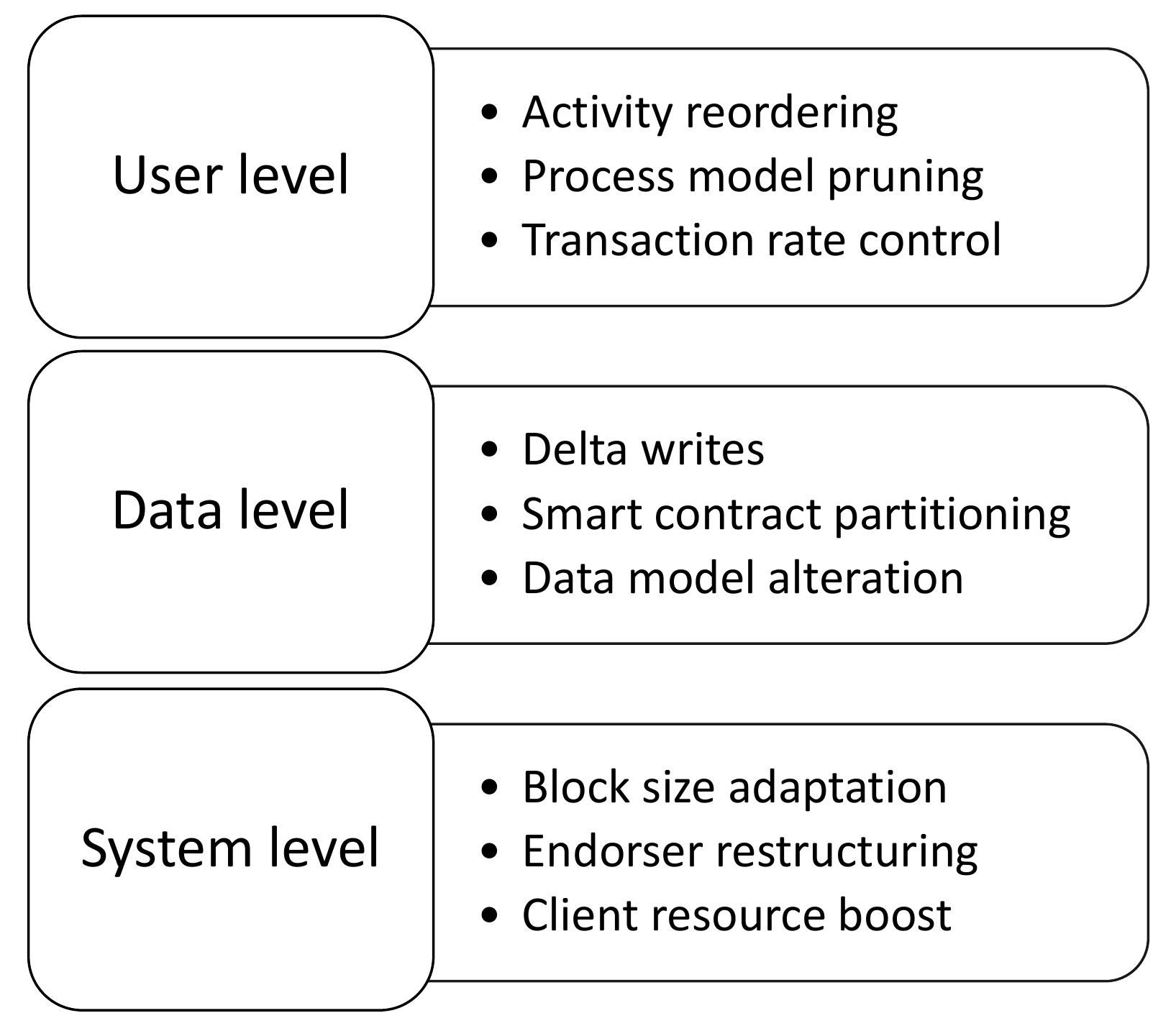}
\caption{Multi-level blockchain optimization}
\label{blockoptrlevels}
\end{figure}
Further, the literature falls short for an end-to-end optimization approach that includes not only system-level tuning and data remodeling but also process model redesign. Since permissioned blockchains are mainly employed by enterprises, a model-driven approach is often followed where the setup of the blockchain network, its transaction regulations, the underlying smart contract, and the data model are primarily based on a business process model created specifically for a particular application~\cite{DBLP:conf/bpm/TranLW18, article, di2019blockchain, 8445060, labazova2021managing}. Such process models may be designed by business domain experts who are unaware of performance implications. For example, in Hyperledger Fabric (a.k.a. Fabric) ~\cite{Androulaki:2018:HFD:3190508.3190538}, many transaction failures arise due to the order in which the transactions are executed~\cite{10.1145/3448016.3452823, 10.1145/3318464.3389693, Sharma:2019:BLB:3299869.3319883}. Such failures could be reduced if the client processes that issue the transactions followed a different business logic in the first place. The prominence of data management while executing business processes has often been highlighted by the database community ~\cite{ 10.1145/2463664.2467796, 10.1145/1989284.1989286, 10.1007/978-3-540-88873-4_17}. We make a similar argument for the importance of the process view in blockchains since the aspects covered by blockchains are manifold and not limited to data alone.

Therefore, given the numerous optimizations possible within a given blockchain system, their varying influence on a case-by-case basis~\cite{10.1145/3448016.3452823, 8526892, 10.1145/3417310.3431398, XU2021102436, Nasir2018PerformanceAO, 8525394}, and the resulting implementation efforts, there is a pressing need for a recommendation system that guides the user in selecting effective optimization strategies suitable for the blockchain under consideration depending on the specific use-case. Again, we can draw parallels from the exhaustive literature on parameter tuning and indexing recommendations for databases~\cite{7495648, ameri2016challenges, 839397, 10.14778/3352063.3352112}. However, since blockchains juggle multiple factors such as organizational governance~\cite{articleorggov}, database definitions~\cite{9143893}, consensus algorithms~\cite{8123011}, provenance tracking~\cite{101555}, and smart contract design~\cite{8494045}, a holistic perspective to optimization recommendations is desirable, which is currently lacking.

To address this gap, we propose a multi-level optimization recommendation approach for blockchains that provides to the users a comprehensive understanding of the different optimization possibilities for their blockchain system, thus enabling them to make a well-informed decision. Inspired by the abstraction levels in databases~\cite{ 10.1145/800227.806871}, we define three levels of abstraction for blockchain optimizations: system, data, and user-level (cf. Figure~\ref{blockoptrlevels}). The system-level recommendations include identifying ideal system configurations such as the block size or endorsement policy. The data-level recommendations deal with understanding the data model and optimizing smart contracts. The user-level recommendations focus on business process models and workloads induced by client processes. For example, we identified two activities in a digital rights management scenario (cf. Section 5.2) that frequently cause transaction conflicts and recommend a process model redesign to reduce such failures. Our approach can also verify compliance with the new process model. We design and implement a recommendation tool named \textsf{BlockOptR} that analyzes the blockchain logs from Fabric, one of the most widely used blockchains by enterprises~\cite{rauchs20192nd}, to demonstrate the performance improvements yielded by our approach.

Our contributions can be summarized as follows:
\begin{enumerate} [nosep, wide, labelwidth=!, labelindent=0pt]
\item We define a multi-level optimization recommendation approach that extensively analyzes the blockchain log and recommends optimization possibilities from different perspectives. Our method helps users gain a comprehensive understanding of their current system and make educated decisions regarding optimization strategies.
\item We provide a formal definition for our recommendation strategies based on common attributes, such that any blockchain log with similar attributes can reuse our approach. We also discuss how our approach translates to different blockchain platforms, thereby providing the reader with a technology-independent outlook.
\item We automate the extraction, preprocessing, and event log generation techniques for Fabric blockchain data. Thus, our tool \textsf{BlockOptR} will help to ease further research in the area of log-based analysis such as process mining in blockchains, since a preprocessed blockchain log can be directly obtained.
\item We demonstrate the effectiveness of the optimization recommendations by implementing and evaluating them. Our experiments indicate an average of 20\% improvement in the percentage of successful transactions and an average of 40\% improvement in latency after applying the recommendations by \textsf{BlockOptR}.
\item We extensively evaluate \textsf{BlockOptR} with three different types of workloads: A set of 24 synthetic workloads generated with a wide range of control variables, four widespread use case-based workloads from the literature, and a real-world event log of a loan application process. Thus, we cover a wide range of scenarios in our experimentation that are representative for real blockchain applications. This aids in overcoming the lack of publicly available data that restricts current research on process mining in permissioned blockchains. The \textsf{BlockOptR} tool, all the smart contracts, the workload generation scripts, and all the event logs are released as open-source to encourage further research in this area~\cite{blockprom}.
\item We further establish the positive effect of our holistic recommendation approach on top of existing blockchain optimizations. Thus, we highlight that \textsf{BlockOptR} complements existing system-level blockchain optimization strategies such as FabricSharp~\cite{10.1145/3318464.3389693} and Fabric++~\cite{Sharma:2019:BLB:3299869.3319883} by adding higher-level optimizations.
\end{enumerate}

\section{Background}
\label{sec:background}

% \begin{figure*}[ht]
% \minipage{0.45\linewidth}
% \includegraphics[width=\linewidth]{figures/scmpmandtrace_new.pdf}
% \captionsetup{justification=centering} 
% \captionsetup{labelfont={bf}}
% \caption{Derived process model for supply chain management scenario}
% %\caption{Effect of the \mbox{number} of organizations}
% \label{scmpmandtrace}
% \endminipage\hfill
% \minipage{0.5\linewidth}
% \includegraphics[width=\linewidth]{figures/scmreorderedpm_new.pdf}
% \captionsetup{justification=centering} 
% \captionsetup{labelfont={bf}}
% \caption{Derived process model for supply chain management scenario after activity reordering}
% \label{scmreorderedpm}
% \endminipage\hfill
% \end{figure*}

\subsection{Hyperledger Fabric}\label{fabricbackground}
Fabric is an open-source permissioned blockchain system popularly used by enterprises~\cite{Androulaki:2018:HFD:3190508.3190538}. The main components of Fabric are a smart contract (called chaincode), a distributed immutable ledger, a distributed world state database, a set of distributed peers, and an ordering service. The smart contract defines all the supported transactions on the blockchain. The transaction flow in Fabric follows an execute-order-validate (EOV) model~\cite{fabricflow}. The EOV model of Fabric is comparable to optimistic concurrency control in databases~\cite{HARDER1984111} and is therefore prone to multi-version concurrency control (MVCC) conflicts, which result in transaction failures. 

\begin{enumerate}[wide, labelwidth=!, labelindent=0pt, nosep]
\item In the execution phase, transaction proposals are created by clients and sent to the endorsers. Endorsers are a set of specific peers that have the authority to execute the smart contract to endorse a transaction. An endorsement policy is configured to define the number of required endorsers for a transaction to be deemed valid. Endorsers generate read-write sets after smart contract execution. The transaction proposal and the read-write sets are signed by the endorsers and sent back to the clients.

\item In the ordering phase, the clients forward these transactions to the ordering service. The ordering service orders the transactions into blocks using Raft~\cite{Ongaro:2014:SUC:2643634.2643666}, a crash fault-tolerant consensus algorithm, and sends them to all the peers in the network. Configurable parameters limit the number of transactions included in a block (block size) in terms of the number of transactions (block count), a timeout (block timeout), or the size of transactions in bytes (block bytes). Blocks are created whenever the buffered set of incoming transactions satisfies any of the three conditions.

\item In the validation phase, every peer validates every transaction. Every peer in the Fabric network has a copy of the distributed ledger and the world state. Peers validate both the endorser signatures based on the endorsement policy and the read-write set with the current world state. If the validation is successful, the world state is updated. Else, a failure is detected. If the endorsement validation fails, it is called an endorsement policy failure; if the read-write set validation fails, it is called an MVCC read conflict. MVCC read conflicts for range reads are called phantom read conflicts. Regardless of the success or failure of the validation, all transactions are appended to the distributed ledger. Also, in the literature, MVCC read conflicts are often classified into inter-block and intra-block failures depending on whether the conflicting transactions reside in the same block or different blocks in the blockchain~\cite{ Sharma:2019:BLB:3299869.3319883,10.1145/3448016.3452823}.
\end{enumerate}

\subsection{Event Logs and Process Mining}\label{pmbackground}
An event log is a record of process executions over time. Process mining~\cite{van2012process} is the technique of deriving a process model that exhibits the most frequent behaviors in an event log. It is mainly used for \emph{process discovery} which helps to understand the underlying process model, \emph{conformance checking} where deviations between a predicted process model and the actual behavior of the process can be identified and \emph{model enhancement} where bottlenecks are identified and removed. The minimum data required in an event log for process mining are:
\begin{enumerate}[nosep]
\item \texttt{CaseID}: To distinguish different executions of the same process. Example: \texttt{ProductID} in a supply chain management related event log. A complete execution of a process is called a trace.
\item \texttt{Activity name}: To identify the different steps in a process. Example: \texttt{Ship} or \texttt{Unload} activity in a supply chain management related event log.
\item \texttt{Timestamp}: To determine the order of the different activities.
\end{enumerate}
The event log can also have other attributes such as process owner, resources, and cost. Various algorithms are used to derive the process model such as alpha miner~\cite{1316839}, heuristics miner~\cite{f8cf09f3d2e243e8afdccaf7ea57ffd8} and fuzzy miner~\cite{gunther2007fuzzy}. The core concept of all these algorithms is to analyze the different traces of the set of activities in the log and simplify the traces through abstraction or aggregation to produce a complete process model. Various open-source and commercial process mining tools are available (ProM~\cite{van2005prom}, Disco~\cite{gunther2012disco}, Celonis~\cite{celonis}).

\section{A Process Perspective to Blockchains}
\label{sec:processperp}

Our work posits blockchain optimization as a holistic methodology rather than a pure system-level approach by introducing a process perspective. In this section, we emphasize the necessity and effectiveness of understanding the dependency between business processes and the performance of the blockchain through examples. Further, these examples motivate the need for an optimization recommender since many process-level optimizations can only be employed with approval from the decision-making bodies of an organization and, in most cases, cannot be automatically applied.
\begin{figure}[ht]
\centering
\includegraphics[width=0.9\columnwidth]{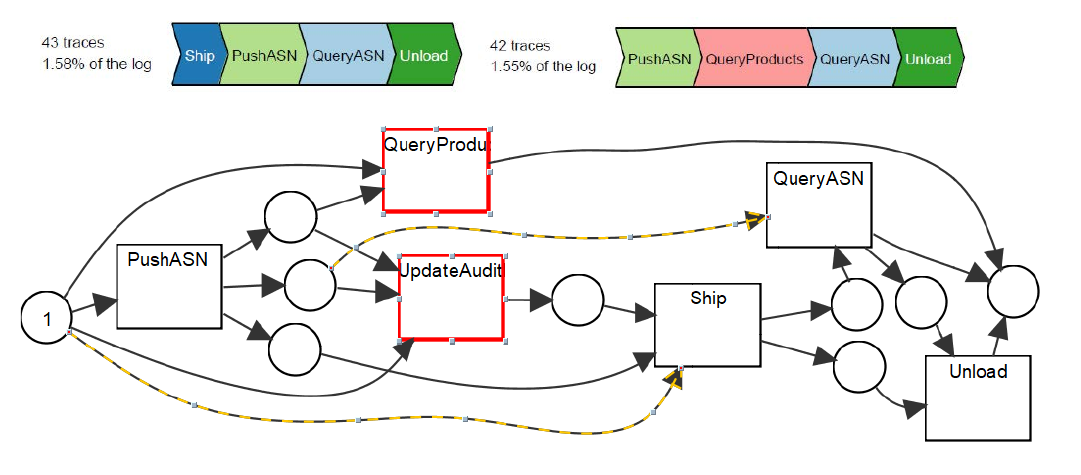}
\caption{Derived process model for SCM scenario}
\label{scmpmandtrace}
\end{figure}

Process model pruning is an example of a process-level optimization that positively affects the system's performance. Figure~\ref{scmpmandtrace} shows the process model derived from the blockchain log of a supply chain management (SCM) scenario. The highlighted paths and the traces embedded in the figure identify two unnecessary branches in the process model. Unless the advanced shipping notice is pushed (PushASN), one should never execute the Ship activity. Similarly, the Unload activity should never be executed unless a product has been shipped. Such illogical activity paths can arise due to manual errors or transaction failures, and the smart contract is designed to handle such issues, as we explain in the following example.

If the Unload transaction executes without a corresponding Ship, the transaction will only read the state but not modify it. However, it is up to the smart contract designer to either fail the transaction upon execution or commit the read-only transaction to the blockchain. Both these designs have their trade-offs. Committing the transaction adds an immutable record on the blockchain, which helps to track, for example, individuals or organizations who deviated from the expected process model. In a supply chain management scenario specifically, this is critical since the entire pipeline is distributed, and the primary purpose of the blockchain here is to provide data provenance among untrusted participants. However, on the other hand, failing a transaction immediately upon execution ensures that such unnecessary transactions do not go through all the time-consuming phases (ordering and validation), which can improve the system performance. We observe a 27\% improvement in throughput and 19\% increase in success rate of transactions when unnecessary activity paths are pruned in the smart contract (Section~\ref{scmresults}, Figure~\ref{scmgraph}). The pruning can also be implemented at the process execution level by enforcing incentive or penalty measures for organizations or individuals that adhere to or deviate from the expected process model. This approach ensures that system performance is not prioritized over data provenance and hence, combines the advantages of both smart contract designs we discussed above.

\begin{figure}[ht]
\centering
\includegraphics[width=0.8\columnwidth]{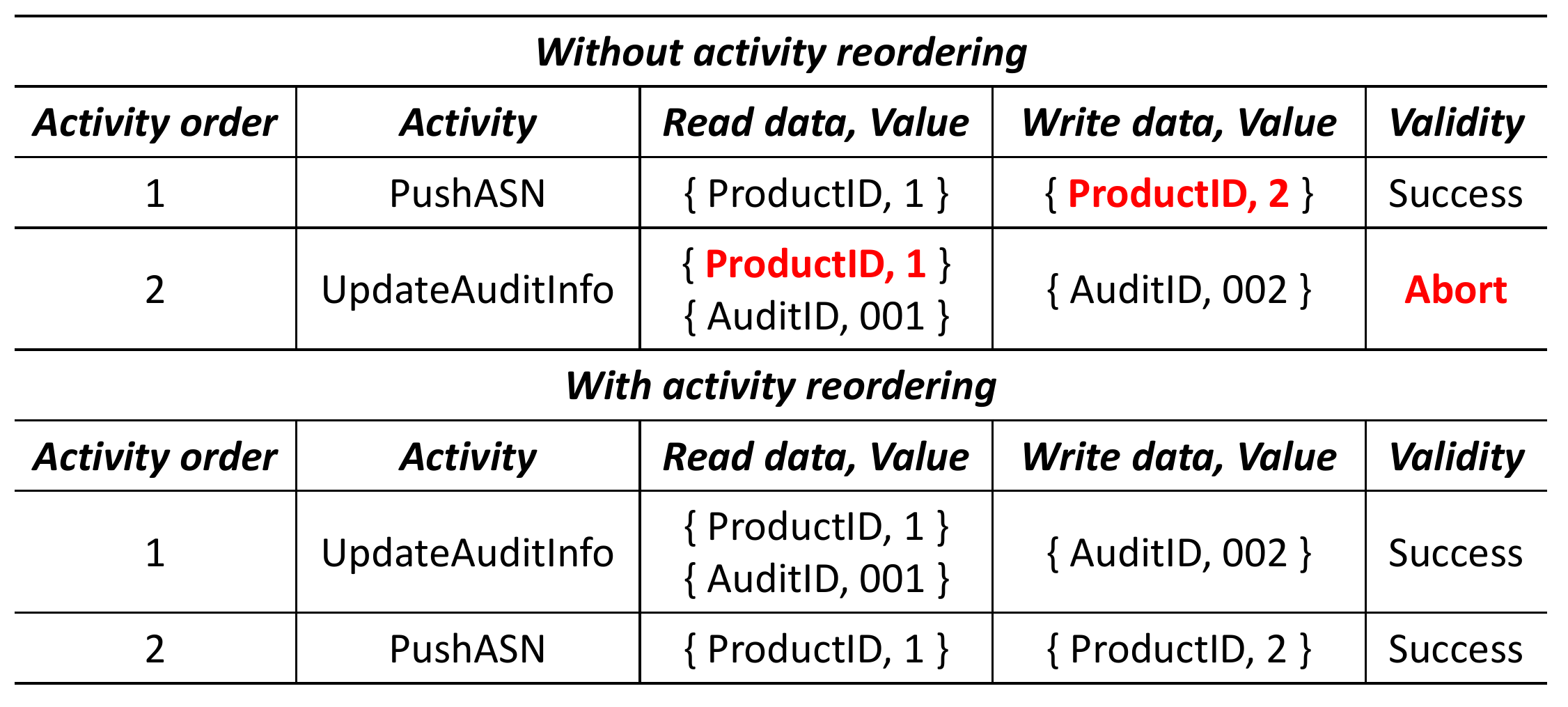}
\caption{Transaction dependency conflict example}
\label{activityreord}
\end{figure}

Another cause of failures are transactional dependencies, and research in serialization algorithms has effectively reduced such failures through transaction reordering~\cite{Sharma:2019:BLB:3299869.3319883, 10.1145/3318464.3389693}. However, reordering algorithms are expensive, as they basically need to solve the NP-hard problem of generating conflict-free dependency graphs~\cite{Sharma:2019:BLB:3299869.3319883}. An increase in endorsement policy failures due to inconsistent world states and the inability to handle large range queries are known problems of transaction reordering~\cite{10.1145/3448016.3452823}. A different approach to the problem of dependency conflicts is to identify  \textsf{reorderable} and \textsf{unreorderable}~\cite{10.1145/3318464.3389693} \emph{activities} instead of transactions. While the literature analyzes the keys accessed by transactions to understand serializability, the data model needs to be analyzed for process-level serialization. If two concurrent activities read the same data element but write to different elements in the data model then such activities are reorderable.

\begin{figure}[ht]
\centering
\includegraphics[width=\columnwidth]{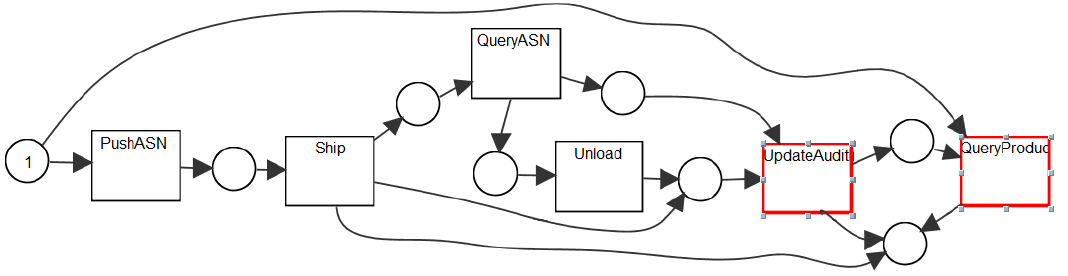}
\caption{Derived process model after activity reordering}
\label{scmreorderedpm}
\end{figure}
For example, in the same supply chain management scenario, the \textsf{UpdateAuditInfo} activity reads a \textsf{productID} and writes an \textsf{auditID}, whereas the 
\textsf{PushASN},  \textsf{Ship}, and  \textsf{Unload} activities read and write to the \textsf{productID}. Therefore, the pairs \{\textsf{UpdateAuditInfo}, 
\textsf{PushASN}\}, \{\textsf{UpdateAuditInfo}, \textsf{Ship}\} and \{\textsf{UpdateAuditInfo}, \textsf{Unload}\} are reorderable activities while \{\textsf{PushASN}, \textsf{Ship}, \textsf{Unload}\} are unreorderable. Figure~\ref{activityreord} shows an example of a reorderable pair of activities where \textsf{UpdateAuditInfo} can succeed if it is executed either after the commit or before the execution of \textsf{PushASN}. Based on the business logic, it may be possible to impose procedures to restrict or reschedule certain activities to execute only at specific periods. For example, the corresponding process model in Figure~\ref{scmpmandtrace} shows that \textsf{UpdateAuditInfo} occurs frequently between \textsf{PushASN} and \textsf{Ship} activities and therefore, \textsf{UpdateAuditInfo} may be executed before the transactions of the other two activities commit. However, \textsf{UpdateAuditInfo} is not a time-critical activity and can be rescheduled to take place only at specific times when traffic is low on the supply chain. We observe a 24\% increase in throughput and 15\% increase in success rate of transactions after a corresponding redesign where \textsf{UpdateAuditInfo} and  \textsf{QueryProducts} activities are executed after { \textsf{PushASN}, 
\textsf{Ship},  \textsf{Unload}}. The new process model derived from the blockchain log confirms the adherence to the new design (Figure~\ref{scmreorderedpm}). Thus, by identifying conflicting activities, the process model can be redesigned to reduce transaction conflicts before they take place.

%It is important to note that this approach only ensures a reduction in the conflicts and not a complete removal of all conflicts, as is the case for transaction-level reordering strategies. This is because the activity execution order may not be maintained by the transaction ordering logic implemented by the blockchain.   

\section{Blockchain Optimization Recommender}
\label{sec:blockprom}

We introduce an approach to recommend optimizations from three different abstraction levels: system, data, and user-level. The primary requirement to design and implement such a multi-level recommendation system is reliable data on all three levels. Knowledge about the system configurations (e.g., block size) and performance (e.g., throughput, transaction failures) is vital for generating system-level recommendations. Information about the current data model and access patterns, such as key distribution and dependencies, is essential for data-level recommendations. Lastly, knowledge concerning the use-case, business processes, and transaction workload is necessary for user-level recommendations. It is important to note that such information is not restricted to a specific level but is helpful across all levels. For example, the system-level performance can indicate the need for optimizations at all three levels.

The very definition of a blockchain implies the availability of a distributed ledger with immutable data regarding every transaction executed overtime. If we consider smart contracts, then every execution of the smart contract results in a transaction that is logged in the ledger. We consider this data (hereafter referred to as the blockchain log) as the primary source to derive optimization recommendations since, to our knowledge, such a distributed ledger consisting of all transactions is available for most blockchains. Therefore, our transaction-centric approach to deriving blockchain optimization recommendations is applicable to different blockchains.

\begin{figure}[t]
\centering
\includegraphics[width=0.75\columnwidth]{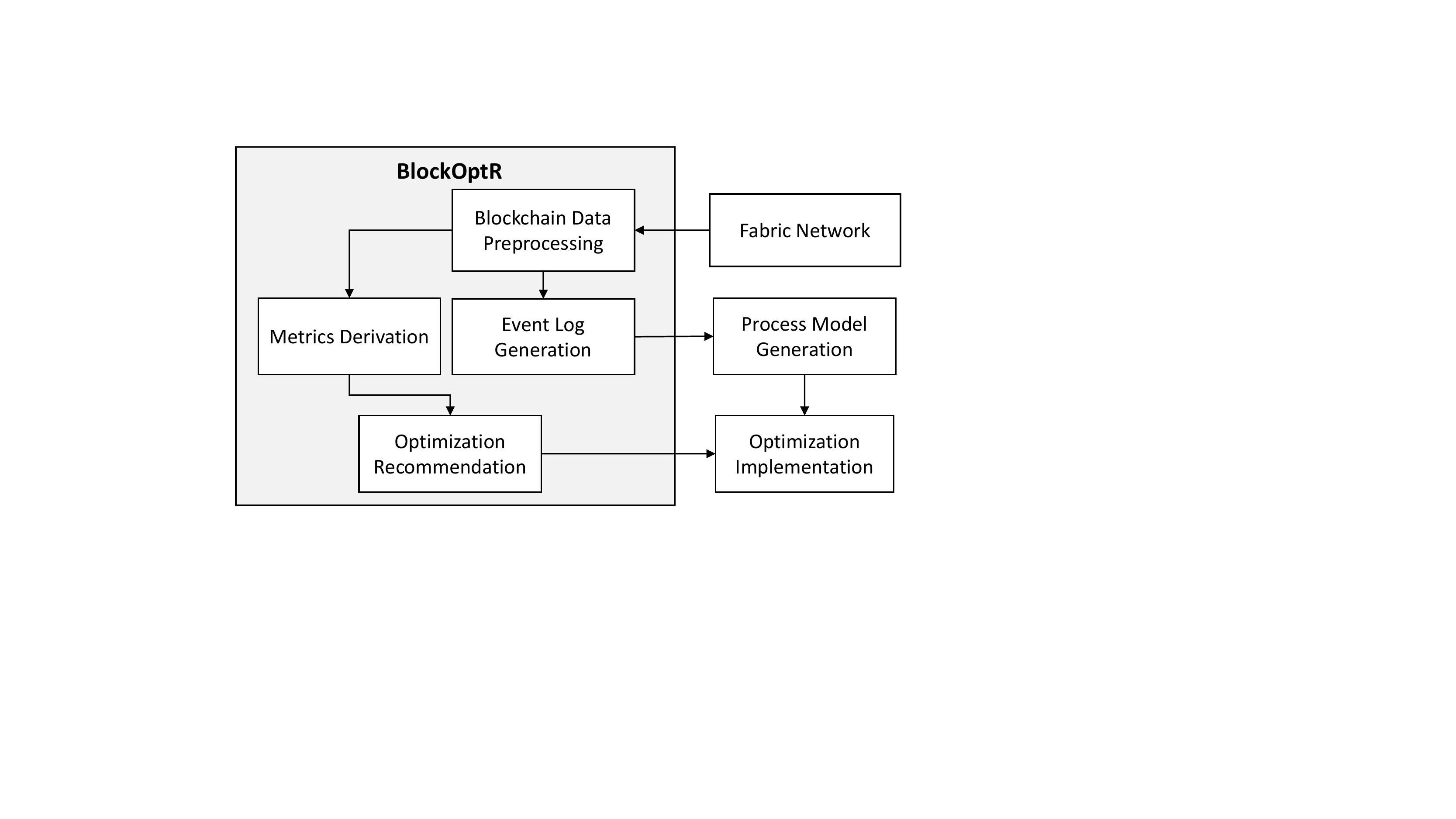}
\caption{BlockOptR workflow}
\label{blockprom}
\end{figure}

We preprocess the raw data from the blockchain to create a blockchain log. Then, we obtain the values for key metrics which are used to detect multi-level optimization recommendations. Process mining strategies are then applied to the blockchain log to derive the process model. We identify the applicable optimizations using the recommendations and the derived process model. Figure~\ref{blockprom} illustrates the workflow of our approach. We automated the main elements of this workflow as a tool, \textsf{BlockOptR}~\cite{blockprom}, implemented in Python and Node.js.

\subsection{Blockchain Data Preprocessing}
\textsf{BlockOptR} registers as a client on the Fabric network, reads the entire blockchain and saves it as JSON files. Next, the log is cleaned by removing the configuration and setup-related transactions that are not relevant and converted to CSV format. All information regarding each transaction executed in the Fabric network is logged on the blockchain. We extract seven attributes and derive two attributes from this extensive logged data. These attributes enable the derivation of multiple metrics required to recommend optimizations. The output of the data preprocessing step is a blockchain log with the following nine attributes.
\begin{enumerate}[nosep]
    \item \textbf{Client timestamp}: The time at which the client generated the transaction.
    \item \textbf{Activity name}: The name of the smart contract function whose execution created the transaction. A(x) defines the activity name of a transaction x.
    \item \textbf{Function arguments}: The value of the parameters of the smart contract function.
    \item \textbf{Endorsers}: The set of all endorsers of the transaction. 
    \item \textbf{Invokers}: The set of all clients and their respective organization who invoked the transactions.
    \item \textbf{Read-write set}: The set of keys accessed (read or written) by the transaction. The separate read set and write set of a transaction are also kept. RWS(x), RS(x) and WS(x) correspondingly define the read-write set, read set and write set of a transaction x.
    \item \textbf{Transaction status}: The status of the transaction that can have the values \texttt{success}, \texttt{MVCC read conflict} (MRC), \texttt{phantom read conflict} and \texttt{endorsement policy failure}. ST(x) defines the status of a transaction x.
    \item \textbf{Transaction type}: The type of transaction which is derived from the read-write set. This can have the values \texttt{read}, \texttt{write}, \texttt{update}, \texttt{range read} and \texttt{delete}. Transaction type is derived from the read-write set. TT(x) defines the type of a transaction x. 
    \item \textbf{Commit order}: The order of the transactions in the blockchain log is the order in which transactions were committed to the blockchain. 

\end{enumerate}

\subsection{Event Log Generation}\label{caseidsec}
The blockchain log extracted from the Fabric network can be used as an event log to apply process mining techniques that assist in recommending user-level optimizations. However, unlike the event logs created by process-aware information systems~\cite{articleaalst}, a \texttt{CaseID} is not directly available in the event log extracted from a blockchain. Also, in most of the use-cases we observed, no single attribute is common to all activities that can be directly used as the \texttt{CaseID}. Therefore, we need to derive a common element for each use-case based on domain knowledge~\cite{caseiddoc, caseiddoc2, casestudy, app112210556, inproceedings}. Since we are interested in a transactional perspective of the process model, we find a common element for all activities by analyzing the function arguments and read-write sets available in the log. For example, in the SCM scenario the \textsf{productKey} is a common element for all activities and is a suitable choice since the use-case is specifically related to tracking multiple products. This process of extracting the common element is automated for all the use-cases in this paper and can be easily extended for other use-cases. Once a common element is identified, we define a trace as a unique set of activities with the same value for the common element. We then assign a new \texttt{CaseID} to every trace.  

Further, only the time at which the clients sent the transaction (client timestamp) is available in our log. However, there is no guarantee that the same order in which clients send their transactions will be maintained when the transactions are committed to the blockchain. Therefore, to derive the process model accurately, we use the commit order in place of the timestamp. Thus, with the generated \texttt{CaseID} and extracted/derived attributes, we have a complete event log. Now, any process mining technique can be applied to the event log to derive a process model. For example, we used the Alpha algorithm to derive the process models shown in Figure~\ref{scmpmandtrace} and~\ref{scmreorderedpm} ~\cite{1316839}.

\subsection{Metrics}
We define a set of metrics by scrutinizing multiple blockchain logs and analyzing metrics from the literature. 

\begin{enumerate}[nosep, wide, labelwidth=!, labelindent=0pt]
\item \textbf{Rate metrics}: \textsf{BlockOptR} calculates the average transaction rate as well as the transaction rate distribution over time intervals from the event log. \textbf{Transaction rate ($\mathit{Tr}$)} is the average rate at which transactions are sent from the clients and is derived from the total transactions in the log and the client timestamps. \textbf{Transaction rate distribution ($\mathit{Trd_i}$)} is the transaction rate at a specific interval $\mathit{i}$ derived from the log. A user-configurable interval size ($\mathit{ins}$) in seconds is used to calculate this metric. \textit{Usage:} Transaction rate is a useful metric to understand the performance. The rate distribution provides insights regarding periods of high or low traffic.
 
\item \textbf{Failure metrics}: Similar to $\mathit{Tr}$, the \textbf{total failure rate ($\mathit{TFr}$)} as well as the rates of each type of failure (MVCC read conflicts, phantom read conflicts, endorsement policy failures) are derived from the log. The failure rate distribution ($\mathit{Frd_i}$) is calculated similar to $\mathit{Trd}$. \textit{Usage:} Failure metrics help to detect times of high transaction failures. Optimizations such as transaction rate control can be applied based on the failure metrics.
 
\item \textbf{Block size}: The user-configured \textbf{block count ($\mathit{B_{count}}$)} and \textbf{block timeout ($\mathit{B_{timeout}}$)} are extracted from the log. The \textbf{average number of transactions in a block ($\mathit{B_{sizeavg}}$}) is also derived from the log. $\mathit{B_{sizeavg}}$ is equivalent to the average block size and can also be defined as $\mathit{min\{B_{count}, Tr* B_{timeout}\}}$.  
\textit{Usage:} $\mathit{B_{sizeavg}}$ along with the rate metrics helps a user understand the effectiveness of their block size configurations. For example, if $\mathit{Tr}$ is 500,  $\mathit{B_{count}}$ is 100, $\mathit{B_{timeout}}$ is 1 and $\mathit{B_{sizeavg}}$ is 100, then 100 transactions are packed into a block when 500 transactions are actually available every second. This means more blocks than necessary are being created which is inefficient, as block creation is expensive. Similarly, if $\mathit{Tr}$ is 100, $\mathit{B_{count}}$ is 500, $\mathit{B_{timeout}}$ is 2, and $\mathit{B_{sizeavg}}$ is 200, then blocks are created only every 2 seconds and transactions are queued up for a waiting period before being put into blocks. Both scenarios lead to performance degradation. So, based on the value of $\mathit{B_{sizeavg}}$, the user can update $\mathit{B_{count}}$ and $\mathit{B_{timeout}}$ to efficiently handle the transaction rate. 
 
% \item \textbf{Block size}: The average number of transactions in every block is recorded as the \textbf{actual block size ($\mathit{ABs}$)}. The \textbf{user-configured block size ($\mathit{Bs}$)} is also extracted from the blockchain. \textit{Usage:} The actual block size used in the Fabric network may differ from the configured one due to block timeouts. Understanding and comparing the real block size and configured block size will help to decide on a better configuration. For example, if we observe that the actual block size is less than the configured block size, we know that the ordering service times out waiting for sufficient transactions to fill the block which can lead to high latency. In this case, a smaller block size or smaller timeouts would be more appropriate. This metric can also be used along with the rate metrics to find a suitable block size that matches the transaction rate as described in the literature~\cite{10.1145/3448016.3452823, 8526892}. 

\item \textbf{Endorser significance ($\mathit{EDsig}$)} defines the number of transactions endorsed by each endorsing peer. \textit{Usage:} This metric helps in identifying endorser bottlenecks. Suppose a limited number of endorsers always carry out the endorsements. In that case, the user can consider distributing the transactions more evenly among the endorsers or expanding the set of endorsers.
 
\item \textbf{Invoker significance ($\mathit{IVsig}$)} defines the number of transactions invoked by each client. \textit{Usage:} This metric helps to identify clients and the corresponding organizations that invoke a majority of the transactions. Client resource allocation decisions of such organizations can be made based on this metric. 

\item \textbf{Key frequency ($\mathit{Kfreq}$)} is defined as the number of failed transactions that access a specific key. \textbf{Key significance ($\mathit{Ksig}$)} is defined as the number of activities that access a specific key. $\mathit{HK}$ defines the set of hotkeys that have high key frequency based on user-configurable thresholds. \textit{Usage:} Identifying the hotkeys assists the users to identify optimization possibilities in their smart contracts, and key significance helps to detect the exact activities (that correspond to smart contract functions) that access the hotkeys. For example, if several functions access the same key, then the different functions could be separated into multiple smart contracts. Every smart contract executes on a different world state, thereby reducing failures (see example in Section~\ref{hotkeysec}).   

 \item \textbf{Data-value correlation ($\mathit{corDV}$)} defines that two transactions are correlated if both access a same set of keys and one of them fails. \textit{Usage:} Data-value correlation helps to identify transaction dependencies. Such dependent transactions are the root cause of MVCC read conflicts~\cite{10.1145/3448016.3452823}. Various optimization strategies, such as process model redesign and transaction rate control, can be applied to these correlated transactions to mitigate failures. 
 \newcommand{\quotes}[1]{``#1''}

 \item \textbf{Proximity correlation ($\mathit{corP}$)} defines the distance between two transactions that have a high data value correlation. For example, if $\mathit{corP(x,y) ==  1}$ then transaction $\mathit{y}$ happened immediately after $\mathit{x}$ with no transactions in between. Further, we also derive the \textbf{activity-based proximity correlation ($\mathit{corPA}$)} which defines the distance between transactions of the same activity. \textit{Usage:} Analyzing if the proximity correlation is \quotes{less than the block size} or \quotes{greater than the block size} can reveal useful insights regarding inter- and intra-block failures.  If intra-block failures are very high, smaller block sizes can potentially reduce failures~\cite{10.1145/3448016.3452823}. This metric also helps to choose between inter- or intra-block transaction reordering strategies offered by different Fabric optimizations~\cite{Sharma:2019:BLB:3299869.3319883,10.1145/3318464.3389693}. 

\end{enumerate}

\subsection{Optimization Recommendations}
We use a multi-level approach to utilize the defined attributes and metrics for recommending blockchain-specific optimization strategies. The optimization recommendation techniques explained in this section include configurable thresholds. We define appropriate default values for these thresholds based on our analysis of multiple logs, but the user can adapt these default values to fine-tune the detection strategies. The necessary condition to recommend each optimization strategy is formalized in Table~\ref{optrecsconditions}.
\subsubsection{\textbf{User Level Recommendations}}
~\\At the user level, it is essential to focus on the actual workload of the running application. The rate and order in which the transactions are generated and committed to the blockchain has a vital impact on performance. We analyze the rate, dependencies, and type of the transactions to recommend optimizations at the user level.

\begin{enumerate}[nosep, wide, labelwidth=!, labelindent=0pt]

\item \textbf{Activity reordering}: Reorderable pairs of transactions can be identified by using the data value correlation and the read-write set. \textsf{BlockOptR} identifies the activities corresponding to such transaction pairs and recommends a process model redesign. The redesign should ensure that the identified activities are restructured to reduce conflicts (cf. Section~\ref{sec:processperp}).

\item \textbf{Process model pruning}: If activities deviate from an expected behavior, then process model pruning is recommended. The transaction type of all transactions related to an activity is analyzed to identify anomalies. Comparing the traces in the event log and the derived process model with the identified anomalies helps to detect model pruning opportunities (cf. Section~\ref{sec:processperp}). 

\item \textbf{Transaction rate control}: \textsf{BlockOptR} evaluates the transaction rate distribution over time and identifies times when the rate is very high. It then checks the failure rates in the same time interval. If the failure rate is also very high, rate control is recommended. Two configurable thresholds are used to tune the tolerance level of transaction rate and failures.

\end{enumerate}

\begin{table}[t]
\begin{center}
 \footnotesize
 \caption{Formalization of optimization recommendations}
 \begin{tabular}{ | m{8em} | m{22em}| } 

 \hline
  \makecell{\textbf{Recommendations}} & \makecell{\textbf{Necessary conditions}}  \\ 
%    \makecell{\textbf{Optimization} \\ \textbf{Recommendations}} & \makecell{\textbf{Necessary conditions} \\ ($\mathit{x,y \in TX, e \in E, c \in I, HK_i \in HK}$) \\ (Config thresholds: $\mathit{Rt_1, Rt_2, Bt, Et, It}$ )}  \\ 
  \hhline{|=|=|}
\makecell{Activity \\ reordering} &  \makecell{$\mathit{if \, corDV(x,y) == 1 \, \land \, WS(x) \cap WS(y) == \emptyset}$} \\
 \hline
 \makecell{Process model \\ pruning} &  \makecell{$\mathit{if \, A(x) = A(y) \land TT(x) \neq TT(y)}$} \\
\hline
 \makecell{Transaction rate \\ control} &  \makecell{$\mathit{if \, (Trd_i \geq Rt_1) \, \land \, (Frd_i \geq Trd_i * Rt_2)\, }$} \\
\hline
 \makecell{Delta writes} &  \makecell{$\mathit{if \, corPA(x,y) == 1 \land ST(x)==MRC\,\land}$ \\ $\mathit{\,|WS(x)|==|WS(y)|==1 \land WS(x) \pm 1 == WS(y)}$} \\
 \hline
 \makecell{Smart contract \\ partitioning} &  \makecell{$\mathit{if \, Ksig(HK_i) > 1}$} \\
 \hline
 \makecell{Data model \\ alteration} &  \makecell{$\mathit{if \, (Ksig(HK_i) == 1) \lor (|HK| == 1)}$} \\
\hline
  \makecell{Block size \\ adaptation} & \makecell{$\mathit{if\, (Tr \geq B_{sizeavg} * Bt) \lor (Tr < B_{sizeavg} * Bt)}$ } \\
  %\makecell{Block size \\ adaptation} & \makecell{$\mathit{if\, (Tr \geq Bs * Bt) \lor (Tr < Bs * Bt)}$ \\ $\mathit{\lor (ABs < Bs * Bt)}$ }  \\ 
  \hline
  \makecell{Endorser \\ restructuring} & \makecell{$\mathit{if \, EDsig(e) > |TX| * Et}$}   \\
 \hline
 \makecell{Client resource\\ boost} &  \makecell{$\mathit{if \, IVsig(c) > |TX| * It}$} \\
  \hline
  \multicolumn{2}{|c|}{}\\
  \multicolumn{2}{|c|}{where $\mathit{x,y \in TX, e \in E, c \in I, HK_i \in HK}$}\\
  \multicolumn{2}{|c|}{$\mathit{TX, E, I, HK}$ are set of all transactions, endorsers, invokers and hotkeys}\\
  \multicolumn{2}{|c|}{$\mathit{Rt_1, Rt_2, Bt, Et, It}$ are user configurable thresholds}\\
 % \multicolumn{2}{|c|}{}\\
  \hline
\end{tabular}
\label{optrecsconditions}
\end{center}
\end{table}

\begin{figure*}[ht]
\setlength{\belowcaptionskip}{-10pt}
\centering
\includegraphics[width=0.85\linewidth]{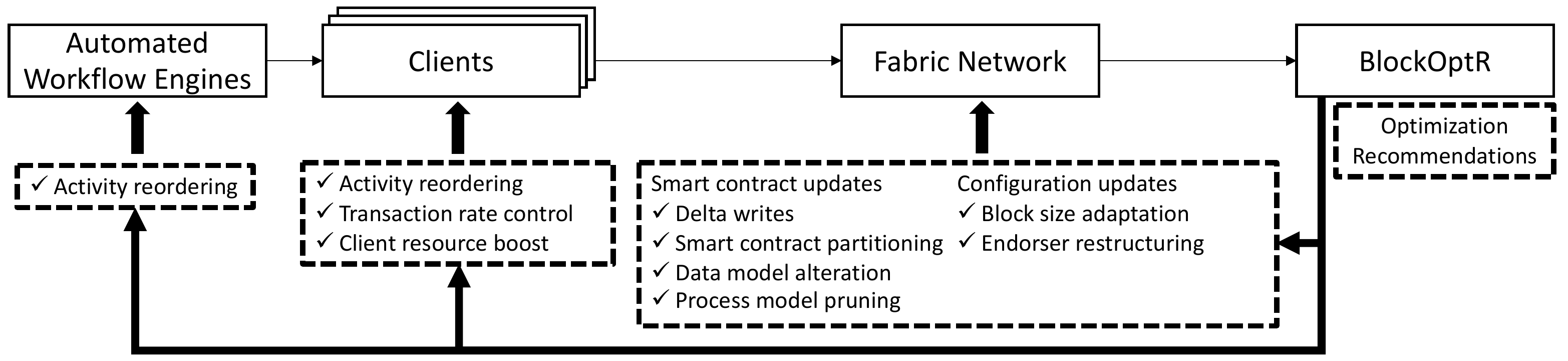}
\caption{Optimization implementation on a live Fabric network}
\label{optimplementation}
\end{figure*}

\subsubsection{\textbf{Data Level Recommendations}}
~\\For data-level recommendations, we focus on identifying the specific areas in the data model that can be optimized by analyzing transaction failures, proximity correlation, read-write sets, and key significance. This aids the user in altering the smart contract and thereby the underlying data model to improve performance.
 
\begin{enumerate}[nosep, wide, labelwidth=!, labelindent=0pt]
\setcounter{enumi}{3}

\item \textbf{Delta writes}: Update transactions that only perform increment or decrement operations can be converted to delta-writes. Delta writes enable writing to multiple unique delta keys, which can be aggregated whenever the current value is required. Reading the key before each write is also not required. Thus, update transactions are converted to write-only transactions that write to unique keys. This helps to reduce transaction dependency-related failures. Delta writes are recommended when a single key is incremented or decremented by a failed transaction.

\item \textbf{Smart contract partitioning}\label{hotkeysec}: A possibility to reduce transaction dependencies is to split a smart contract into multiple ones. Each smart contract will access separate world states, thereby avoiding conflicts. The functionality of the original smart contract will not change because it is possible to invoke functions between the two smart contracts if interaction is required. 

For example, in a music rights management scenario, if a key \texttt{MusicID} is found to be hot and multiple functions such as \texttt{Play()} and \texttt{viewMetaData()} access this same key, then one can separate the functions into two different smart contracts. In other words, the underlying database is split into two by duplicating the primary key (\texttt{MusicID}) across both and having different secondary keys in each. The play count of \texttt{MusicID} is recorded in one and metadata is read from the other (cf. Section~\ref{drmresults}). This is analogous to designing the table layout in relational databases. The smart contract needs to be analyzed and updated to implement this optimization. Smart contract partitioning is recommended if multiple activities access a hotkey.

%For example, if a key \texttt{StudentID} is found to be hot and multiple functions such as \texttt{updateGrade()} and \texttt{updatePersonalInfo()} access this same key, then one can separate the functions into two different smart contracts. In other words, the underlying database is split into two by duplicating the primary key (\texttt{StudentID}) across both and having different secondary keys in each. Grades are recorded in one and personal information in the other. This is analogous to designing the table layout in relational databases. The smart contract needs to be analyzed and updated to implement this optimization. Smart contract partitioning is recommended if multiple activities access a hotkey.

\item \textbf{Data model alteration}: If activities have a dependency on themselves, then a data model alteration can be beneficial to reduce transaction conflicts. For example, in a digital voting scenario, if a key \texttt{ElectionID} is found to be hot and is only accessed by the function \texttt{Vote()}, then a possible optimization is to use another primary key such as \texttt{VoterID}. Then, instead of updating all the votes together, the votes can be updated per voter (cf. Section~\ref{dvresults}). Further, if only a single hotkey is detected then it is beneficial to analyze the data model to understand the reason for the skewed access to this specific data element (cf. Section~\ref{lapresults}). Data model alteration is recommended if a hotkey is accessed only by a single activity or if a single hotkey is detected.

%For example, if a key \texttt{TeacherID} is found to be hot and is only accessed by the function \texttt{updateGrade()}, then a possible optimization is to use another primary key such as \texttt{SubjectID}. Then, instead of updating all the grades together, the grades can be updated per subject. 

\end{enumerate}
\subsubsection{\textbf{System Level Recommendations}}
~\\At the system level, we focus on two crucial system configuration settings that can significantly affect the performance of Fabric: the endorsement policy and the block size~\cite{10.1145/3448016.3452823,8526892}. Further, we also identify client bottlenecks to aid in resource allocation decisions. We use the endorser significance, invoker significance, transaction rate, and actual block size metrics to derive system-level optimization recommendations. Since these recommendations are based on the blockchain log generated by the running application, it helps the user to identify ideal configuration settings based on their current use-case and workload, leading to performance improvements.

\begin{enumerate}[nosep, wide, labelwidth=!, labelindent=0pt]
\setcounter{enumi}{6}
\item \textbf{Block size adaptation}: The average transaction rate ($\mathit{Tr}$), the average block size ($\mathit{B_{sizeavg}}$) and a configurable threshold ($\mathit{Bt}$) are used to recommend block size adaptation. The literature recommends smaller block sizes when transaction rates are lower and larger block sizes when the rates are higher~\cite{10.1145/3448016.3452823, 8526892}. If the block size is too small, too many blocks are created, and block creation becomes a bottleneck. If the block size is too large, the block creation is delayed by waiting for sufficient transactions. Therefore, if block size adaptation is recommended, then set $\mathit{B_{timeout}}$ and $\mathit{B_{count}}$ such that $\mathit{min\{B_{count}, Tr* B_{timeout}\}}$ is equal to $\mathit{Tr}$. We do not provide recommendations for \emph{block bytes} adaptation since it is difficult to accurately derive the size of a transaction (that can include the transaction payload, endorser identities and other metadata) from the log. 
% \item \textbf{Block size adaptation}: The average transaction rate ($\mathit{Tr}$), the user-configured block size ($\mathit{Bs}$) and the actual block size ($\mathit{ABs}$) derived from the log are compared with an upper and lower threshold ($\mathit{Bt1}$, $\mathit{Bt2}$) to detect a possible block size optimization. The literature recommends smaller block sizes when transaction rates are lower and larger block sizes when the rates are higher~\cite{10.1145/3448016.3452823, 8526892}. If the block size is significantly smaller than the transaction rate, too many blocks are created, and block creation becomes a bottleneck. If the block size is significantly larger than the transaction rate, the block creation is delayed by waiting for sufficient transactions (depending on the block timeout). Therefore, we recommend updating the block size to match the transaction rate if it falls above or below the configurable thresholds. Further, if we observe that the actual block size is less than the configured block size, smaller block size or smaller timeouts are recommended. 

\item \textbf{Endorser restructuring}: For every Fabric transaction generated by the clients, the corresponding smart contract function is executed by the endorsers defined in the endorsement policy. Smart contract execution is a time and resource-consuming action. If the same endorsers receive a higher load of transactions while others remain idle, this indicates a bottleneck or load imbalance. Such load imbalances can occur when the endorsement policy explicitly defines an endorsement as mandatory from a specific set of endorsers. For example, the endorsement policy \texttt{And(Org1,OR(Org2,Org3))} implies that an endorsement from \texttt{Org1} is mandatory. As a consequence, \texttt{Org1} could become an endorsement bottleneck. 
We detect endorser bottlenecks by identifying endorsers that endorse more transactions than a user-specified threshold. The default threshold values detect whether all the endorsers participate equally in the endorsement process. The threshold values can be adapted to increase or decrease the sensitivity to imbalances.

\item \textbf{Client resource boost}: Multiple time-consuming tasks are performed by the clients in a Fabric network, including but not limited to transaction proposal invocation, endorser response verification, packing of endorser responses as a transaction, transaction submission to the ordering service, and collection of peer commit responses. The invoker significance metrics identify the clients and the corresponding organizations that invoke a majority of the transactions. This identification can assist in resource allocation decisions, such as increasing the number and size of clients registered to the identified organization. It could also point to problems in the underlying business process.

\end{enumerate}

\subsection{Implementation of Optimizations}
The recommended optimizations can be implemented in several ways. Figure~\ref{optimplementation} visualizes where the different recommendations can be implemented on a live Fabric network. Here, we show an automated workflow engine that triggers transactions based on a process model. These transactions are sent via the clients to the Fabric network. The logs of the Fabric network are analyzed by BlockOptR to generate optimization recommendations. Each of the recommended optimizations can be implemented at different levels as shown in the figure.

Activity reordering can be implemented by modifying the underlying process model in the workflow engine such that activities follow a conflict-free order. Alternatively, one can monitor the transactions on the clients and reorder either per client or across all clients using a client manager. Process model pruning can be implemented via organizational measures such as incentives or penalties to ensure that activities adhere to their expected behavior (not shown in the figure). However, pruning can also be implemented directly in the smart contract by early aborting anomalous transactions during the endorsement phase. Transaction rate control can be implemented in multiple ways. Each client can monitor their own transaction rate and perform load shedding or queuing. The same can be done across clients using a central monitor. A third approach is to monitor the transaction rate in the ordering service and apply load shedding there. Smart contract revisions are required to implement all the data-level optimizations. In Fabric, smart contract upgrades are possible on the fly without restarting the system~\cite{upgradesc}. Block size can be adapted either by changing the configuration file or by using a \textit{configuration update} transaction in Fabric~\cite{configtx}. Endorser restructuring can be implemented by altering the endorsement policy. The policy can be changed in the Fabric configuration file or using a \textit{configuration update} transaction~\cite{configtx}. Based on the transaction load per client identified by \textsf{BlockOptR}, client resources can be scaled if the current allocation appears insufficient to handle the load and the new clients can be dynamically registered to the Fabric network.

%Conflict-free replicated data types (CRDTs) could also be employed to implement delta-writes~\cite{10.1145/3361525.3361540}.
%However, the real-life organizational hierarchy needs to be considered when changing the endorsement policy which would require management's authorization. 

\textbf{Our implementations}. Although all optimizations can be applied in a live system on the fly, since our evaluation runs in an experimental environment, we restart the Fabric network after every experiment. We use the Caliper benchmarking system~\cite{caliper} which has a client manager that can be configured to order the transactions across clients and control the rate of transactions generated, thus emulating activity reordering and transaction rate control. The number of clients can also be scaled to demonstrate a client resource boost. Process model pruning and all data-level optimizations are implemented by analyzing and modifying the smart contract. Block size and endorsement policies are updated in the Fabric configuration file.

\section{Experimental Methodology}
\label{sec:methodology}
We used version 2.0 of HyperledgerLab~\cite{10.1145/3448016.3452823}, which is an automated testbed for Hyperledger Fabric 2.2 integrated with the Caliper benchmarking system. We set up a Kubernetes cluster of 1 master and 5 worker nodes over which all the Fabric network components as well as Caliper components are distributed as Kubernetes pods. Each node runs on a Ubuntu Focal (20.04) virtual machine with 4 vCPUs and 9.8 GB RAM. We use 10 Caliper workers for our experiments. For every experiment, we measure the success rate which is the percentage of successful transactions out of the total number of transactions, the average latency and the throughput of all successful transactions.

% Table generated by Excel2LaTeX from sheet 'CV'
\begin{table}[t]
  \centering
  \caption{Control variables}
  \footnotesize
    \begin{tabular}{|p{10.69em}|p{11.81em}|}
    \hline
    \textbf{Control Variable} & \textbf{Values} \textbf{(Default in bold)} \\ \hhline{|=|=|} 
    Workload type & \textbf{Uniform}, Read-heavy, Insert-heavy, Update-heavy, RangeRead-heavy \\
    \hline
    Endorsement policy & P1, P2, \textbf{P3}, P4 \\
    \hline
    Endorser distribution skew & \textbf{0}, 6 \\
    % \hline
    % Endorser distribution skew & \textbf{0}, 2 \\
    \hline
    Key distribution skew & \textbf{1}, 2 \\
    \hline
    Number of organizations  & \textbf{2}, 4 \\
    \hline
    Block count & 50, \textbf{300}, 1000 \\
    \hline
    Send rate & 50, \textbf{300}, 1000 \\
    \hline
    Transaction dist skew & \textbf{0}, 70\% \\
    \hline
    \end{tabular}%
  \label{cv}%
\end{table}%
\subsection{Workload Generation}
The content of the distributed ledger, which is used as the input to our tool, is a direct result of the workload executed on the blockchain. Therefore, we extensively evaluate \textsf{BlockOptR} by using three different types of workload. Also, after implementing the recommendations generated by \textsf{BlockOptR}, we rerun the experiments with the same workloads to analyze the effect of the optimization.  

\subsubsection{Synthetic workloads}\hfill

We use an extended version of a synthetic workload generator that can generate synthetic workloads based on a set of control variables for a generic smart contract \textit{genChain}~\cite{10.1145/3448016.3452823}. We use a range of values for these control variables described in Table~\ref{cv} to generate multiple workloads of 10,000 transactions each. The endorsement policies used in our experiments are: \\
P1: \texttt{And(Org1, Or(Org2,Org3,Org4))}\\
P2: \texttt{And(Or(Org1,Org2), Or(Org3,Org4))}\\
P3: \texttt{Majority(Org1,...,OrgN)}\\ %where N = Number of orgs.
P4: \texttt{OutOf(2,Org1,Org2,Org3,Org4)}

By generating synthetic workloads, we ensure that multiple realistic scenarios are covered in our experiments. We then evaluate \textsf{BlockOptR} with each of these workloads to generate optimization recommendations. Further, we implement each of the recommended optimizations to evaluate the performance improvement. 

\subsubsection{Use-case based workloads} \hfill

Secondly, we use extended versions of four popular use-case based smart contracts from the literature~\cite{10.1145/3448016.3452823} and generate workloads. \textsf{BlockOptR} is then used to generate optimization recommendations with these workloads. The four smart contracts we use are as follows.

\newcolumntype{P}[1]{>{\centering\arraybackslash}p{#1}}
\newcolumntype{M}[1]{>{\centering\arraybackslash}m{#1}}
\renewcommand{\arraystretch}{0.3}
\renewcommand{\arraystretch}{1.2}
\begin{table}[t]
  \centering
  \caption{Experiments with the synthetic workload}
  \small\addtolength{\tabcolsep}{-5pt}
  \scalebox{0.7}{%
    \begin{tabular}{|c|c|c|c|c|}
    \hline
    \multicolumn{1}{|M{5.5em}|}{\textbf{Experiment Number}} & \multicolumn{1}{M{10em}|}{\textbf{Control variable}} & \multicolumn{1}{M{10em}|}{\textbf{Value}} & \textbf{Optimizations recommended} \\
    \hhline{|=|=|=|=|}
    {1} & \multicolumn{1}{c|}{{Endorsement}} & \multicolumn{1}{c|}{{P1}}    & Endorser restructuring \\
         & Policy      &           & Activity reordering \\
    \hline
    {2} & \multicolumn{1}{c|}{{Endorsement Policy / }} & \multicolumn{1}{c|}{{P2 / 6 }}    & Endorser restructuring \\
          & Endorser dist skew      &          & Activity reordering \\
    \hline
    {3} & \multicolumn{1}{c|}{{No: of orgs }} & {4}     & Transaction rate control \\
    \hline
    {4} & \multicolumn{1}{c|}{{Workload}} & \multicolumn{1}{c|}{{Read-heavy}}     & Activity reordering \\
    \hline
    {5} & \multicolumn{1}{c|}{{Workload}} & \multicolumn{1}{c|}{{Update-heavy}}     & Transaction rate control \\
    \hline
    {6} & \multicolumn{1}{c|}{{Workload}} & \multicolumn{1}{c|}{{Insert-heavy}}     & Activity reordering \\
    \hline
    {7} & \multicolumn{1}{c|}{{Workload}} & \multicolumn{1}{c|}{{RangeRead-heavy}}    & Activity reordering \\
    \hline
    {} & \multicolumn{1}{c|}{{ }} & {}     & Transaction rate control \\
  8        & Key      &           & Activity reordering \\
          &  distribution skew    &  2        & Smart contract partitioning \\
    \hline
    {} & \multicolumn{1}{c|}{{}} & {}     & Block size adaptation \\
  9        & Block count      &  50        & Activity reordering \\
    \hline
     {} & \multicolumn{1}{c|}{{}} & {}     & Transaction rate control \\
 10         & Block count       &   300        & Activity reordering \\
    \hline
     {} & \multicolumn{1}{c|}{{}} & {}     & Transaction rate control \\
  11        &  Block count     &  1000        & Activity reordering \\
    \hline
    12    & \multicolumn{1}{c|}{Send rate} & 50       & Activity reordering \\
    \hline
     13    & \multicolumn{1}{c|}{Send rate} & 300       & Activity reordering \\
    \hline
    {} & \multicolumn{1}{c|}{{}} & {}     & Block size adaptation \\
              &       &          & Transaction rate control \\
   14       &  Send rate     &  1000        & Activity reordering \\
    \hline
    {} & \multicolumn{1}{c|}{{}} & \multicolumn{1}{c|}{{}}    & Transaction rate control \\
           15       &  Transaction     &  70\%        & Activity reordering \\
          & distribution skew    &           & Client resource boost \\
    \hline
    \end{tabular}}
  \label{snyexperiments}
\end{table}%

\emph{Supply Chain Management (SCM)}: This smart contract defines the operations of a logistics network that includes sending an advanced shipping notice, shipping a product, reading the shipping notice and unloading the product (in this order). There is also a query operation to query the information of the different products (\texttt{queryProducts}) and a \texttt{updateAuditInfo} function that updates an audit entry with the product details.  These can happen at any point in time. We generated a workload of 10,000 transactions based on these assumptions by sending in order the transactions \texttt{pushASN}, \texttt{ship}, \texttt{queryASN} and \texttt{unload} while the transactions \texttt{queryProducts} and \texttt{updateAuditInfo} are sent randomly.

\emph{Digital Rights Management (DRM)}: This smart contract manages the rights of artists in the music industry. The smart contract includes a \textit{Play} function that is executed whenever a piece of music is played by any user. The other smart contract functions include adding a new piece of music, querying the rights, viewing the metadata and calculating the revenue of the right holders. In a realistic scenario, the \textit{Play} transaction would be executed far more frequently than all the other transactions. Therefore, we create a \textit{Play} heavy workload for this use-case. We generate 10,000 transactions randomly where 70\% of the transactions are \textit{Play}. The remaining 30\% comprise all the other transactions generated uniformly at random.

\emph{Electronic Health Records (EHR)}: In this smart contract, patients can provide or revoke access rights to medical institutes as well as research institutes to query their medical records. We assume that the number of patients would be more than the other participants and generate a 70\% update-heavy workload of 10,000 transactions.

\emph{Digital Voting (DV)}: This smart contract includes a function to vote in an election, query the parties, query the results as well as end the election. We can assume that during an actual election there will be periods of high traffic while the voting is taking place. Therefore, we generate a workload which initially sends 1,000 \texttt{queryParties} transactions at a rate of 100 TPS, then 5,000 \texttt{Vote} transactions at a rate of 300 TPS and finally 1 \texttt{seeResults} and \texttt{endElection} transaction each.

\subsubsection{Loan Application Process (LAP)} \hfill

Thirdly, we created a smart contract and workload using a real-life event log of the loan application process of a Dutch financial institute which is available publicly~\cite{lapdata} together with the corresponding process flow~\cite{lappm}. We extracted all the events of the first 2,000 loan applications and created 20,000 corresponding transactions. We then created a smart contract where every activity in the loan application process flow has a corresponding smart contract function. The event log contains an \texttt{employeeID} for every employee in the bank handling loan applications and an \texttt{applicationID} for every loan application processed by the bank. The smart contract we implemented uses the \texttt{employeeID} as the key and the value of the key is an array of structures where every structure includes the \texttt{applicationID}, \texttt{loan type}, \texttt{loan amount} and \texttt{loan status}. Therefore, querying a specific \texttt{employeeID} will easily provide all the applications processed by that employee. We then executed the 20,000 transactions on the smart contract at a low rate of 10 TPS to simulate a real world scenario where manually processing the loan applications takes a long time. We also ran the same experiment at a higher rate of 300 TPS to emulate an automated loan application and validation process. We use \textsf{BlockOptR} to generate optimization recommendations which help to improve the smart contract implementation and thereby the performance.

Though the LAP event logs are from a database setting, this is a realistic use-case for blockchains as an automated loan application system requires security and decentralized trust (e.g., micro-loans, decentralized loan applications, and more generally DeFi~\cite{10.1093/jfr/fjaa010, HugoHoffmann+2021+1+15, Qiao2019/12, 10.1145/3327960.3332395}). Consequently, this experiment demonstrates the utility of \textsf{BlockOptR} in a realistic scenario. In the use-case based experiments, all the transactions followed the expected order based on the assumptions we defined. In contrast, with this real event log, we evaluate the real order in which the transactions are executed which can deviate from the process model.

%\begin{comment}
\newcolumntype{L}{>{\centering\arraybackslash}m{5.3cm}}
\begin{table}[t]
    \centering
    \caption{Settings to implement optimization}
     \footnotesize\addtolength{\tabcolsep}{-5pt}
    \begin{tabular}{| c | L |}
    \hline
        \textbf{Optimizations} & \textbf{Settings} \\ 
        \textbf{Recommended} &  \\ \hhline{|=|=|}
        Activity reordering & Reorder workload generation\\ \hline
        Transaction rate control & Set send rate to 100 TPS\\ \hline
        Process model pruning &   \\ \cline{1-1} 
        Delta writes & Update smart contract \\ \cline{1-1} 
        Smart contract partitioning  & \\ \cline{1-1} 
        Data model alteration  & \\ \hline
        Block size adaptation & Set block count to derived transaction rate \\ \hline
        Endorser restructuring & Set endorsement policy to P4\\ \hline
        Client resource boost & Double clients for recommended organization \\ \hline
    \end{tabular}
    \label{optvaribles}
\end{table}
%\end{comment}
%\hhline{|=|=|}

\section{Experimental Results}
\label{sec:results}
\newcommand{\apostrophe}{{\quotefont'}}

\begin{figure*}[ht]
\setlength{\belowcaptionskip}{-10pt}
\minipage{0.3\linewidth}
\includegraphics[width=\linewidth]{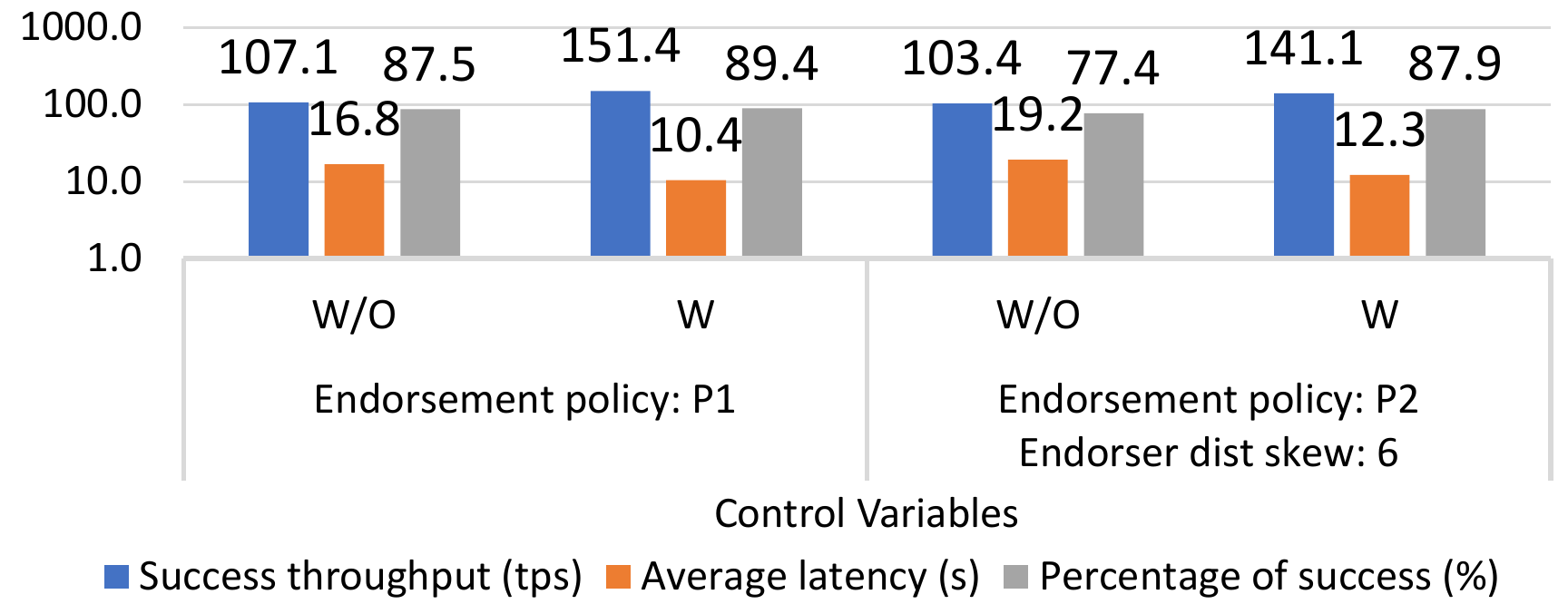}
\captionsetup{justification=centering} 
\captionsetup{labelfont={bf}}
\caption{Endorser restructuring}
%\caption{Effect of the \mbox{number} of organizations}
\label{endpol}
\endminipage
\minipage{0.3\linewidth}
\includegraphics[width=\linewidth]  {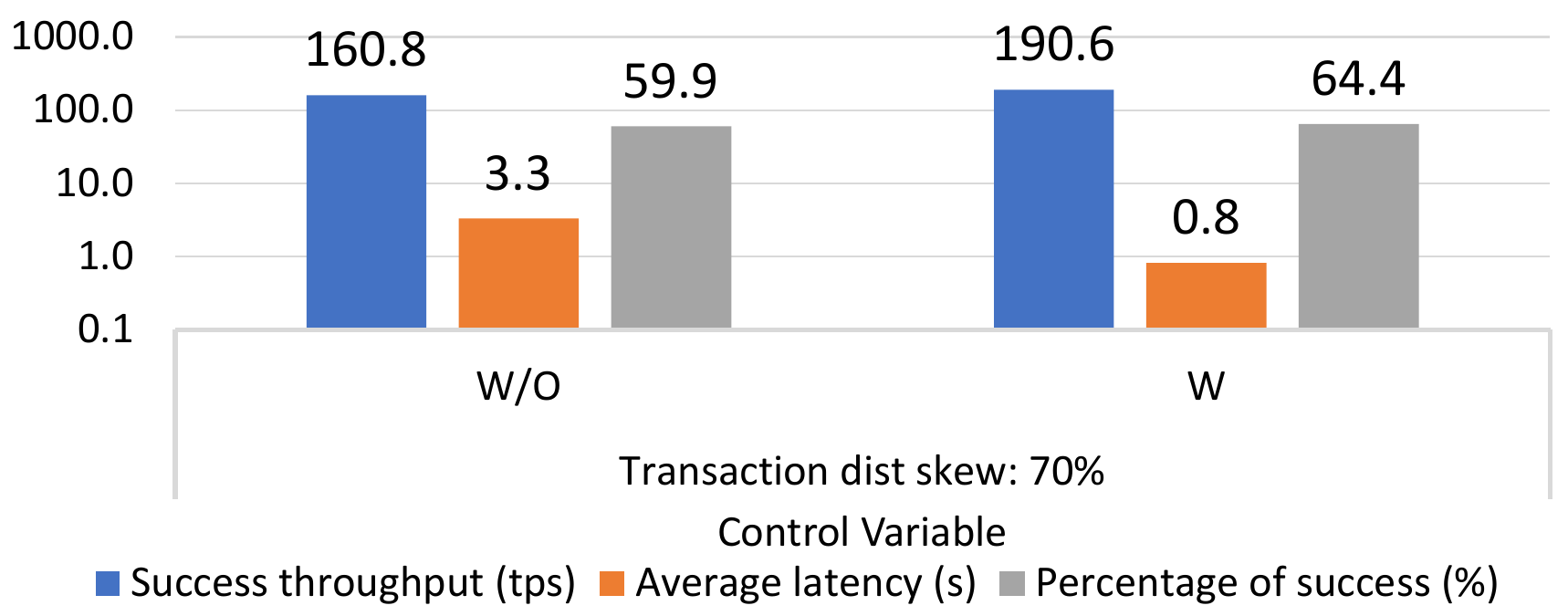}
\captionsetup{justification=centering} 
\captionsetup{labelfont={bf}}
\caption{Client resource boost}
\label{clientdist}
\endminipage
\minipage{0.35\linewidth}
\includegraphics[width=\linewidth]  {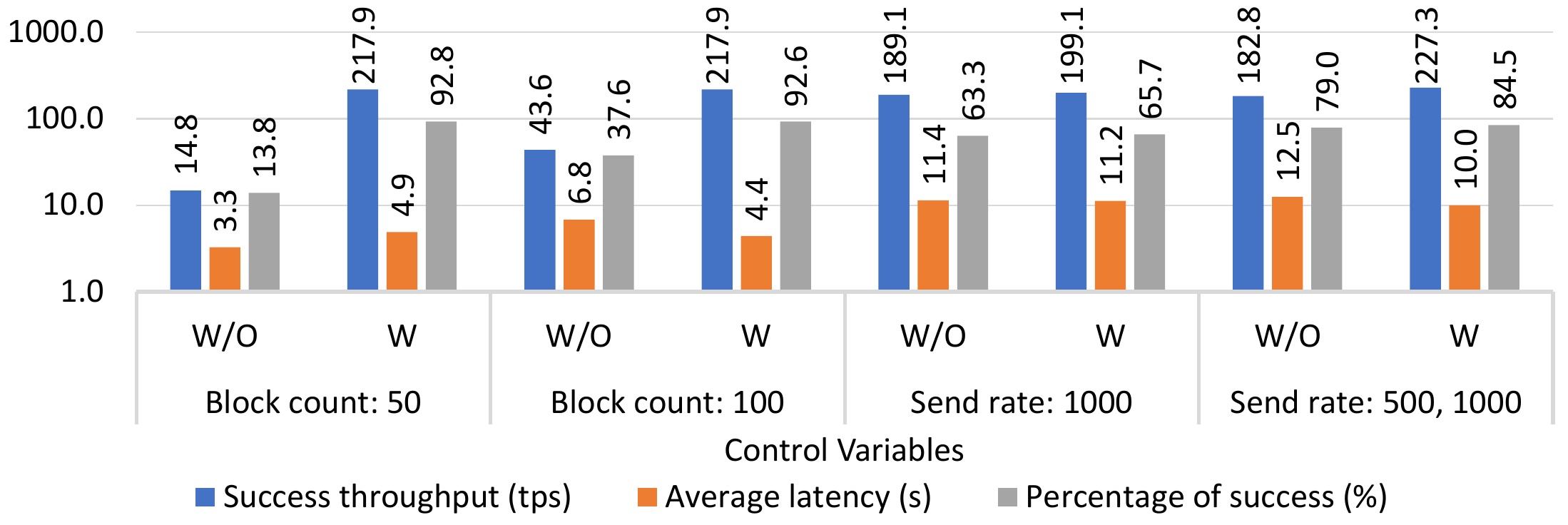}
\captionsetup{justification=centering} 
\captionsetup{labelfont={bf}}
\caption{Block size adaptation}
\label{blksz}
\endminipage
\end{figure*}

% \begin{figure*}[ht]
% \centering
% \includegraphics[width=\columnwidth]{figures/blksz.pdf}
% \caption{Performance without and with block size optimization}
% \label{blksz}
% \end{figure*}

\begin{figure*}[ht]
\setlength{\belowcaptionskip}{-10pt}
\centering
\includegraphics[width=0.95\linewidth]{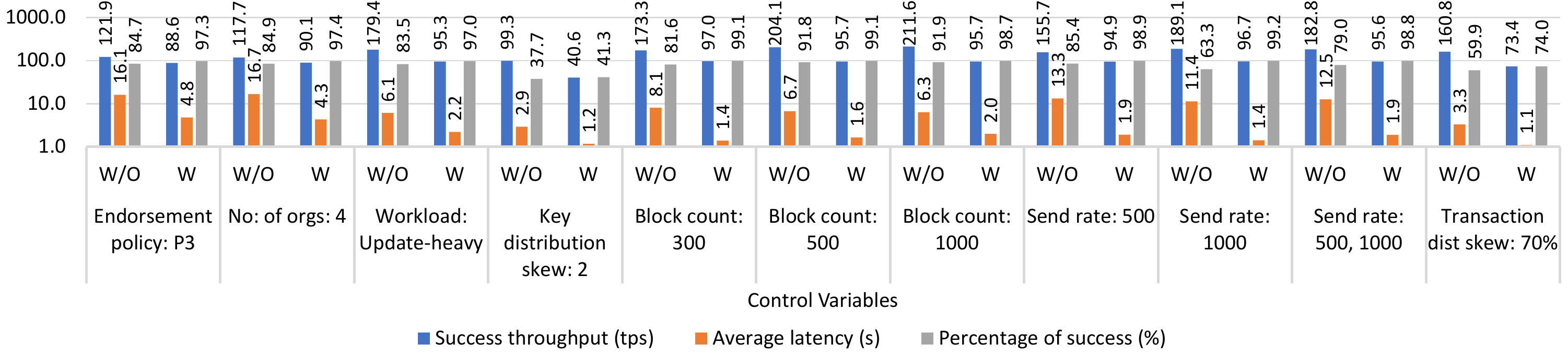}
\caption{Transaction rate control}
\label{ratecntrl}
\end{figure*}

\begin{figure*}[ht]
\setlength{\belowcaptionskip}{-10pt}
\centering
\includegraphics[width=0.95\linewidth]{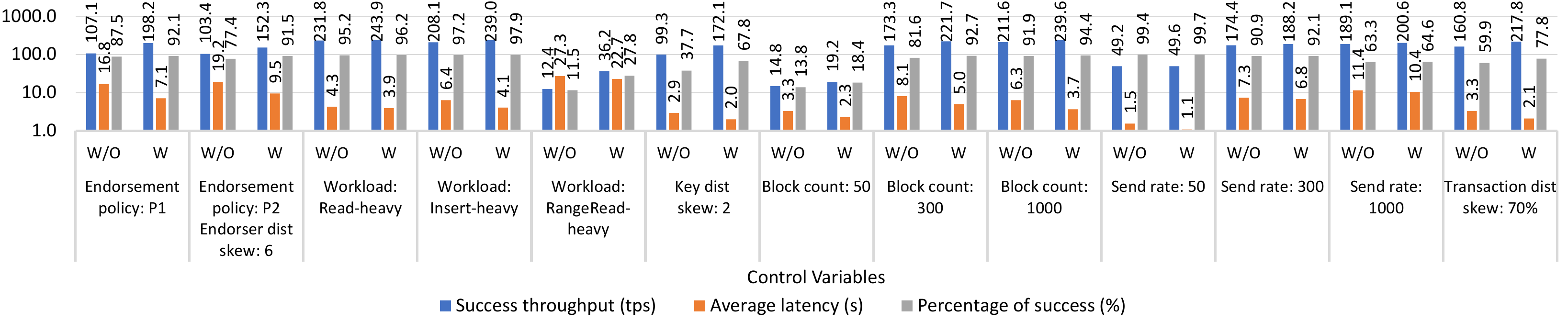}
\caption{Activity reordering}
\label{rord}
\end{figure*}

\begin{figure*}[ht]
\setlength{\belowcaptionskip}{-10pt}
\centering
\includegraphics[width=0.95\linewidth]{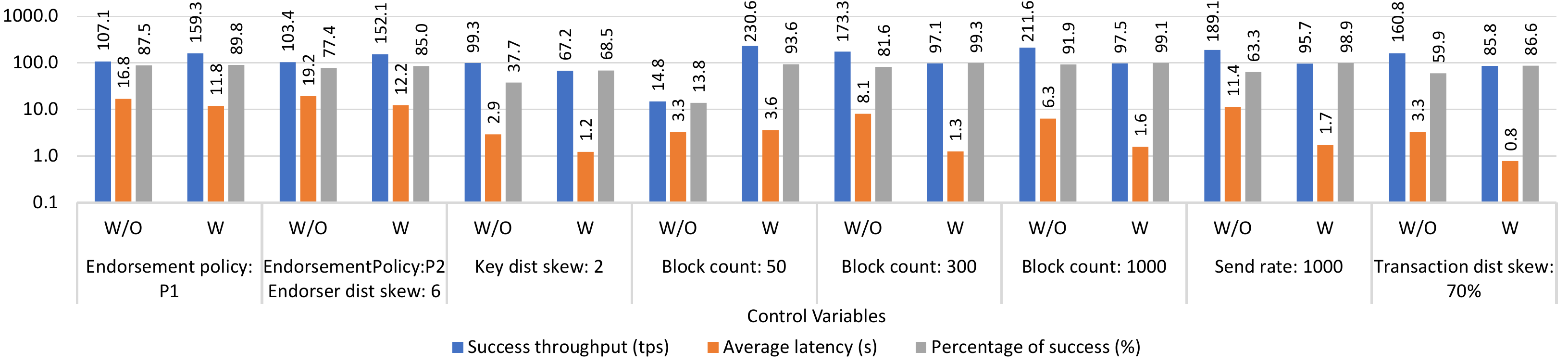}
\caption{All recommended optimizations combined}
\label{allopt}
\end{figure*}

%The primary purpose of the experiments in this section is to 
We exhaustively evaluate our recommendation approach with a wide range of workloads and smart contracts. Please note that, whenever transaction rate control is implemented there is an expected decrease in the throughput. However, clients benefit heavily from higher success rates, and the apparent decrease in the throughput is just closer to the sustainable throughput of the system. In all our experiments the default value for the thresholds are $\mathit{Et=0.5, Rt1=300, Rt2=0.3, Bt=0.6}$ and $\mathit{It=0.5}$. All the settings including the control variable values changed to implement each recommended optimization is shown in Table~\ref{optvaribles}.

%When fine tuning the thresholds and parameters, it should be possible to approach the optimal throughput.

\subsection{Synthetic Workloads}
Due to space restrictions, we present 15 workloads in Table~\ref{snyexperiments}. The full list of experiments and results can be seen in our repository~\cite{blockprom}. The control variable that is tuned for each experiment is shown along with its value. All the other control variables have the default value shown in Table~\ref{cv}. Experiments 1 to 15 are conducted with no optimizations applied and then \textsf{BlockOptR} is used to derive optimization recommendations. The recommendations generated by \textsf{BlockOptR} are also shown in Table~\ref{snyexperiments}. Since the synthetic smart contract has a simple logic with no branches, increment/decrement operations or complex data model, process model pruning, delta writes and data model alterations are not recommended here. Next, we implement the recommended optimizations and re-execute all the experiments. The results of the experiments are grouped based on the optimization recommendations and can be seen in Figures~\ref{endpol}, ~\ref{clientdist},~\ref{blksz},~\ref{ratecntrl},~\ref{rord} and ~\ref{allopt}. We also explain how the thresholds are set for our experiments and how they can be tuned by users.

\begin{figure*}[t]
\setlength{\belowcaptionskip}{-10pt}
\minipage{0.35\linewidth}
\includegraphics[width=\linewidth]{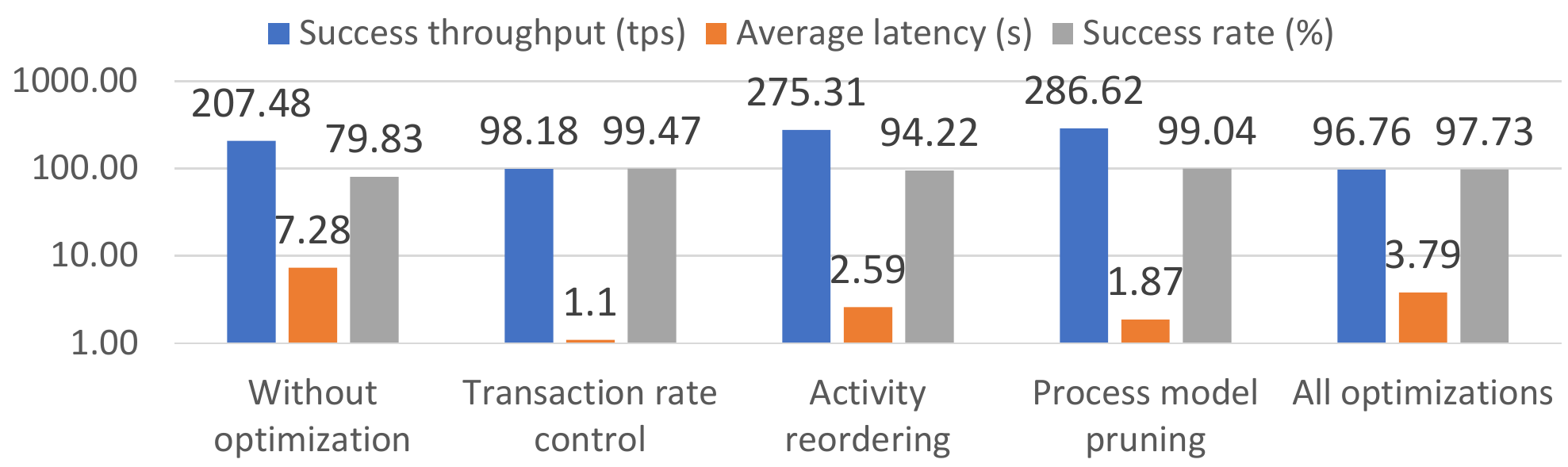}
\captionsetup{justification=centering} 
\captionsetup{labelfont={bf}}
\caption{SCM use-case}
%\caption{Effect of the \mbox{number} of organizations}
\label{scmgraph}
\endminipage
\minipage{0.3\linewidth}
\includegraphics[width=\linewidth]  {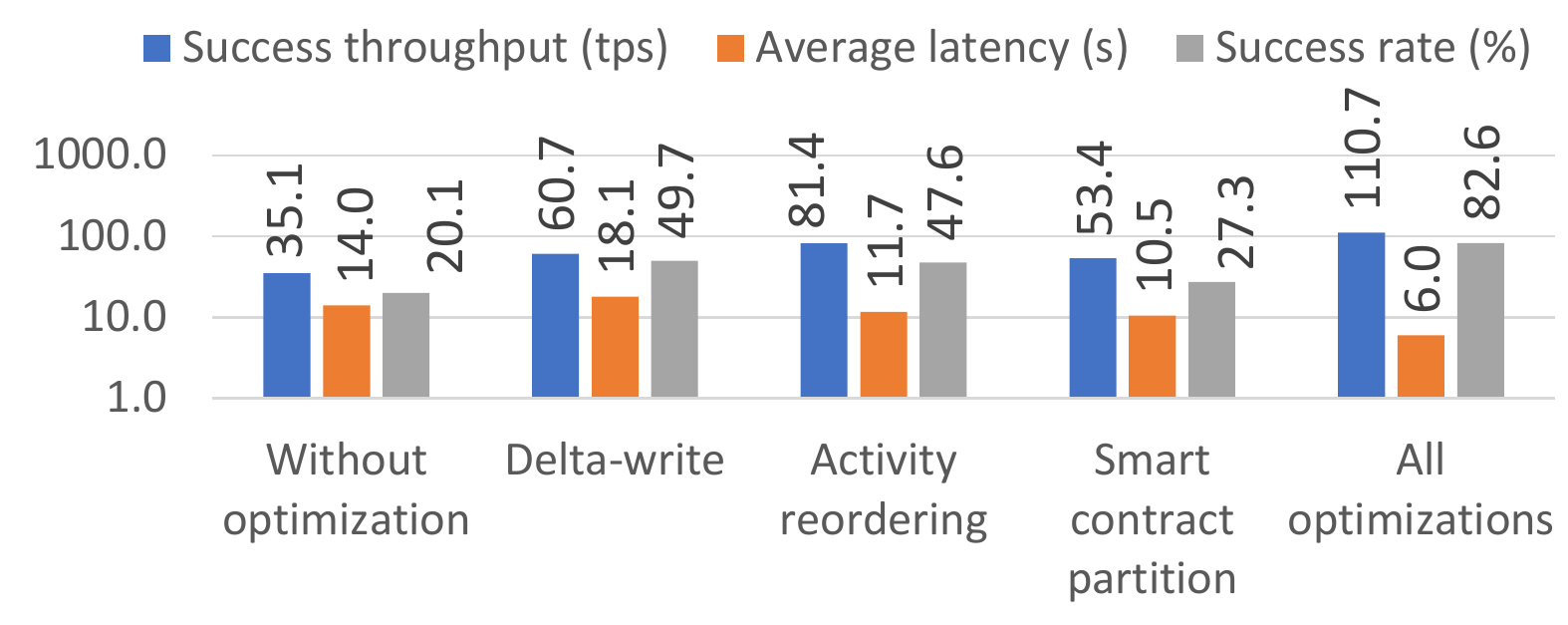}
\captionsetup{justification=centering} 
\captionsetup{labelfont={bf}}
\caption{DRM use-case}
\label{drmgraph}
\endminipage
\minipage{0.3\linewidth}
\includegraphics[width=\linewidth]  {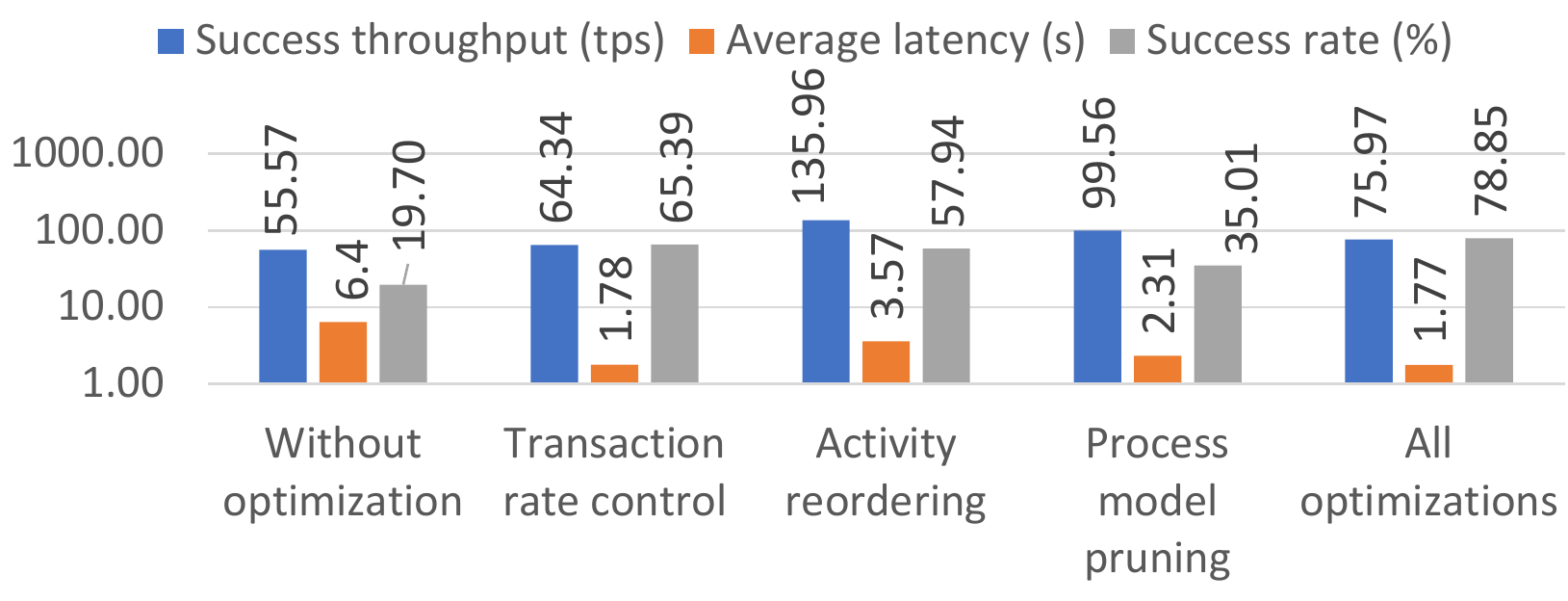}
\captionsetup{justification=centering} 
\captionsetup{labelfont={bf}}
\caption{EHR use-case}
\label{ehrgraph}
\endminipage
\end{figure*}

\begin{figure*}[t]
\setlength{\belowcaptionskip}{-10pt}
\minipage{0.3\linewidth}
\includegraphics[width=\linewidth]{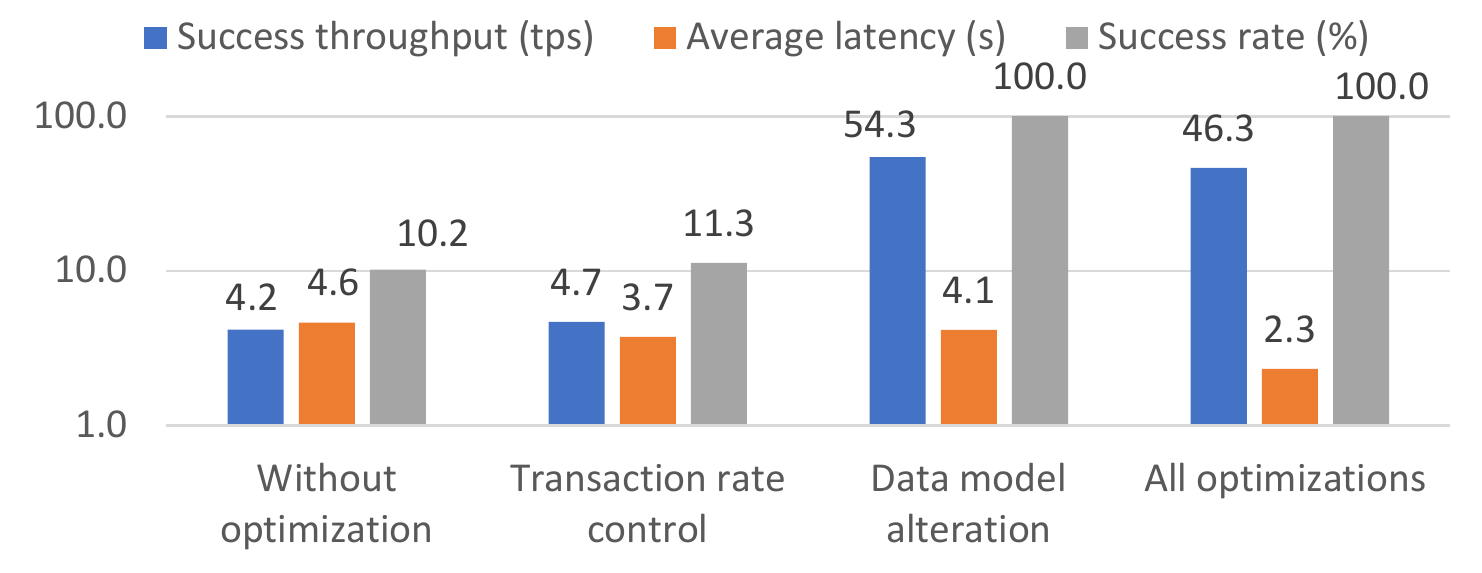}
\captionsetup{justification=centering} 
\captionsetup{labelfont={bf}}
\caption{Digital voting use-case}
%\caption{Effect of the \mbox{number} of organizations}
\label{dvgraph}
\endminipage
\minipage{0.5\linewidth}
\includegraphics[width=\linewidth]  {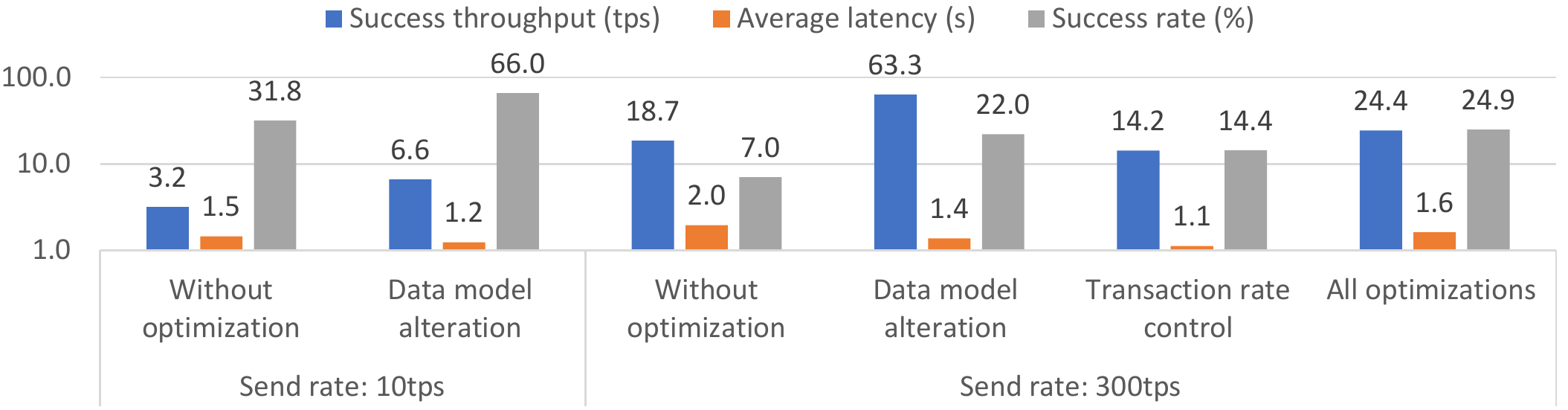}
\captionsetup{justification=centering} 
\captionsetup{labelfont={bf}}
\caption{Loan application process use-case}
\label{lapgraph}
\endminipage
\end{figure*}

\subsubsection{\textbf{Endorser restructuring}}: 
The effect of endorser restructuring can be seen in Figure~\ref{endpol}. When the endorsement policy is P1, all the clients must send their transactions to Org1 due to the specific endorsement policy and hence, an endorsement bottleneck is detected for \texttt{Org1}. Since the endorsement policy requires signatures from two organizations, we change the policy to \texttt{OutOf(2,Org1,Org2,Org3,Org4)} so that the clients can distribute the transactions evenly among all endorsers. This optimization leads to a 29\% increase in throughput (Figure~\ref{endpol}). In Experiment 2, since the endorser distribution is skewed, the clients send transactions unevenly and therefore two of the organizations endorse far more often than the other two. We re-executed the experiment with an even distribution of transactions to the endorsers and observe a 26\% increase in throughput (Figure~\ref{endpol}). The main impact of this optimization is on throughput and latency as it reduces transaction queuing on few specific peers and instead distributes them evenly.    

We set the thresholds for this recommendation such that we expect an even distribution of transactions to all endorsers, i.e., even minor bottlenecks are detected. This can be tuned to detect only severe bottlenecks. Further, since these are synthetic experiments, changing the endorsement policy is not critical. In real scenarios, consultation with the governing bodies of an enterprise is required before changing the policy. Still, the recommendations by \textsf{BlockOptR}  help to highlight bottlenecks  which in turn can convince the management to change the policy. 
\subsubsection{\textbf{Client resource boost}}:
Figure~\ref{clientdist} shows the effect of client resource scaling. After increasing the number of clients, we observe a 75\% decrease in latency, a 15\% increase in throughput, and a 7\% increase in success rate. The thresholds are set such that this optimization is recommended when more than 50\% of transactions are invoked by the same organization. This can be fine-tuned to detect less severe bottlenecks.
\subsubsection{\textbf{Block size adaptation}}:
The effect of block size adaptation can be seen in Figure~\ref{blksz}. In our experiments, we use the default block time out of 1s. Therefore, we make the block count equal to the transaction rate whenever the block size adaptation is recommended. After changing the block size, we observe up to 93\% improvement in throughput and 85\% improvement in success rate (Figure~\ref{blksz}; Block count: 50). The thresholds are set such that this optimization is recommended whenever the average block size is 60\% larger or smaller than the transaction send rate derived from the log. The thresholds can be decreased to make the recommendation more sensitive to transaction rate changes.
% \subsubsection{\textbf{Block size adaptation}}:
% The effect of block size adaptation can be seen in Figure~\ref{blksz}. When the block size is significantly higher or lower than the transaction rate, block size adaptation was recommended. After changing the block size, we observe up to 93\% improvement in throughput and 85\% improvement in success rate (Figure~\ref{blksz}; Block size: 50). The thresholds are set such that this optimization is recommended when the block size is 60\% larger or smaller than the transaction send rate derived from the log. The thresholds can be decreased to make the recommendation more sensitive to transaction rate changes. 
\subsubsection{\textbf{Transaction rate control}}:
The effect of transaction rate control is shown in Figure~\ref{ratecntrl}. In these experiments, periods of high traffic (around 300 TPS) were also identified as periods of high failure rates. We then lowered the transaction send rate to 100 TPS on the clients and re-executed the experiments. We observe significant improvement of up to 87\% in latency and 36\% in success rate (Figure~\ref{ratecntrl}; Send rate: 1000). We set the thresholds for this recommendation at 300 TPS which is the default send rate of our experiments. This means that we consider the current traffic of the system as high and want to detect periods of failure. Users can adjust this threshold based on what is considered high (more than the sustainable traffic rate) for their Fabric network installation. 

\subsubsection{\textbf{Activity reordering}}:
The effect of activity reordering can be seen in Figure~\ref{rord}. We observe that \textsf{BlockOptR} recommends activity reordering for all experiments except Experiments 3 and 5 (Table~\ref{snyexperiments}). Reordering was suggested for two activities (\texttt{Read} and \texttt{Update}) which conflict with each other. We updated the configuration of the client manager to generate read transactions before all other transactions. This implementation emulates a scenario where organizational measures were applied to enforce activity reordering. We then re-executed the experiments and observe a performance improvement in all the experiments. There is up to 65\% increase in throughput and 58\% increase in success rate (Figure~\ref{rord}; Workload: RangeReadheavy). We have set the thresholds such that if 40\% of the MVCC failures are caused by activities that can be reordered, this strategy is recommended. This can be made more lenient by increasing the threshold such that reordering is suggested only in very significant cases. For Experiments 3 and 5, less than 40\% of MVCC conflicts are caused by the two activities where reordering is possible. For example, the activity \texttt{Update} has a dependency on itself which cannot be removed by reordering.
%It is important to note that activity reordering only ensures a reduction in the conflicts and not a complete removal of all conflicts, as is the case for transaction-level reordering strategies~\cite{Sharma:2019:BLB:3299869.3319883, 10.1145/3318464.3389693}. This is because the activity execution order may not be maintained by the transaction ordering logic implemented by the blockchain.

\subsubsection{\textbf{Combined optimizations}}:
We also executed the experiments after applying all the recommended optimizations together. We observe up to a 93\% improvement in throughput and 85\% improvement in the success rate (Figure~\ref{allopt}: Block count: 50). In all the experiments, the performance obtained by applying all the optimizations is comparable to the performance yielded by the optimization with the highest improvement. 

%This shows that multiple optimization strategies may not be required in most scenarios and a single optimization can already yield comparable improvements in performance. For example, activity reordering is recommended for almost all the experiments, which significantly improves the performance and leaves little room for further improvements. However, there are scenarios when reordering is not a viable strategy. For example, if there is a strict process model that cannot be changed, reordering the activities will not be viable. Therefore, \textsf{BlockOptR} recommends multiple optimizations from which the user can select on a case-to-case basis. Further, the user can also apply each optimization, generate new blocks and then rerun \textsf{BlockOptR} to decide whether further optimizations are necessary. 

\textbf{Further remarks}. Though smart contract partitioning is recommended for Experiment 8, this optimization requires understanding the functionality of the smart contract. Unfortunately, for the synthetically generated smart contract that includes only generic read, update and insert functions, we cannot redesign the smart contract.

\subsection{Use-case based Workloads \label{usecasebased}} 

% \begin{figure}[ht]
% \centering
% \includegraphics[width=\columnwidth]{figures/scmgraph.pdf}
% \caption{Supply chain management use-case}
% \label{scmgraph}
% \end{figure}

\emph{Supply Chain Management (SCM)}\label{scmresults}:
With the SCM use-case, three optimizations are recommended by \textsf{BlockOptR}: activity reordering, process model pruning and transaction rate control (Figure~\ref{scmgraph}). After implementing reordering for the reorderable activities (\texttt{query\-Products} and \texttt{UpdateAuditInfo}), we observe a 24\% increase in throughput and 15\% increase in success rate. Pruning was recommended for the \texttt{Ship} activities that occur without or before the \texttt{PushASN} activity. It was also recommended to prune \texttt{Unload} activities that occur without or before the \texttt{Ship} activity. We adapted the smart contract to implement the pruning recommendation. This resulted in a 27\% improvement in throughput and 19\% increase in success rate. Transaction rate control and applying all recommendations together also improves the performance.

% \begin{figure}[ht]
% \centering
% \includegraphics[width=0.9\columnwidth]{figures/drmgraph.pdf}
% \caption{Digital rights management use-case}
% \label{drmgraph}
% \end{figure}
\emph{Digital Rights Management (DRM)}\label{drmresults}: 
With the DRM use-case, three optimizations are recommended by \textsf{BlockOptR}: activity reordering, delta-writes and smart contract partitioning. Figure~\ref{drmgraph} shows the results of applying these optimizations. To implement the delta write recommendation, we observed that the \texttt{Play} function in the smart contract has an increment operation to count the number of times a piece of music was played. We converted this into a delta write and the delta-keys are aggregated whenever the \texttt{calcRevenue} function is invoked (since it requires the play count). With this optimization, we can observe a significant improvement of 42\% in throughput and 50\% in success rate. However, the average latency increases in this case because the \texttt{calcRevenue} function now takes up more time for aggregation. Since \texttt{calcRevenue} is not executed as frequently as \texttt{Play}, the overall performance is not affected though. 

Activity reordering was recommended for \texttt{calcRevenue} and \texttt{queryRightHolders} functions and we reconfigured the clients to send these activities after all other activities. This emulates a scenario where an organization restricts specific transactions to specific time periods. We observe more than 50\% increase in both throughput and success rate with this optimization. 

Hot keys were detected and frequently used by four activities. We analysed the smart contract and discovered that, though all four functions have a dependency on the same key, the functionalities are different. \texttt{Play} and \texttt{calcRevenue} need only the play count, while \texttt{viewMetaData} and \texttt{queryRightHolders} need metadata and not the play count of a piece of music. Therefore, we split the smart contract into two, where one smart contract has the \texttt{Play} and \texttt{calcRevenue} functions and the second smart contract has the other two functions. The \texttt{create} function is included in both smart contracts, and invocation of the first smart contract invokes the same function in the second smart contract. We observe a 35\% increase in throughput and a 26\% increase in success rate with this optimization. Applying all the optimizations together improves the performance by more than 50\%. 
%However, when we derived the process model, we observed that the \texttt{calcRevenue} activity was missing which led to the discovery that the corresponding transactions were timing out. With the delta write implementation, \texttt{calcRevenue} now requires more time and with the reordering optimization, these high-latency transactions are being sent together. This led to a large number of timeouts for \texttt{calcRevenue}. Increasing the timeout setting resolved this issue. This example showcases the adverse effects of applying the recommended optimizations without understanding their impact on each other. 

%\begin{figure*}[t]
%\setlength{\belowcaptionskip}{-10pt}
%\minipage{0.4\linewidth}
%\includegraphics[width=\linewidth]  {figures/fabricsharp_generator.pdf}
%\captionsetup{justification=centering} 
%\captionsetup{labelfont={bf}}
%\caption{Synthetic workloads with FabricSharp}
%\label{fsgen}
%\endminipage
%\minipage{0.6\linewidth}
%\includegraphics[width=\linewidth]{figures/fabricplusplus_generator.pdf}
%\captionsetup{justification=centering} 
%\captionsetup{labelfont={bf}}
%\caption{Synthetic workloads with Fabric++}
%\label{fpp}
%\endminipage
%\end{figure*}

\begin{figure}[t]
\includegraphics[width=0.9\columnwidth]  {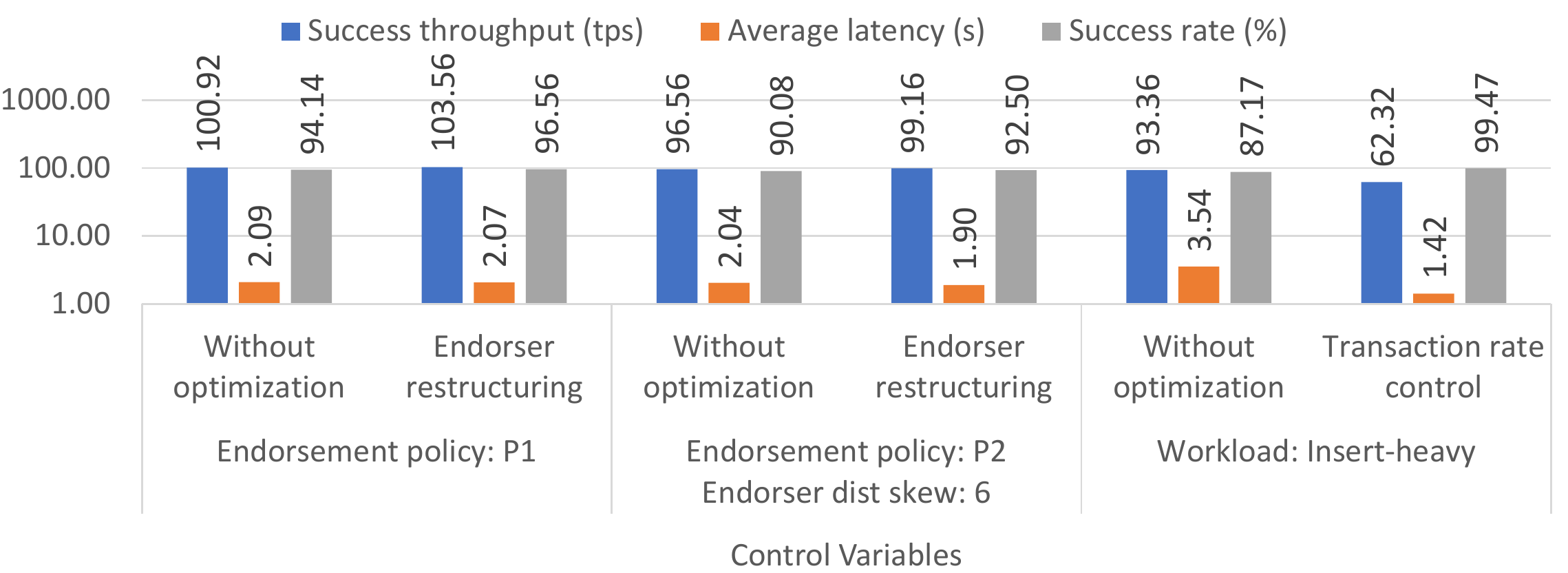}
\captionsetup{justification=centering} 
\captionsetup{labelfont={bf}}
\caption{Synthetic workloads with FabricSharp}
\label{fsgen}
\end{figure}

\begin{figure*}[t]
\includegraphics[width=0.6\linewidth]{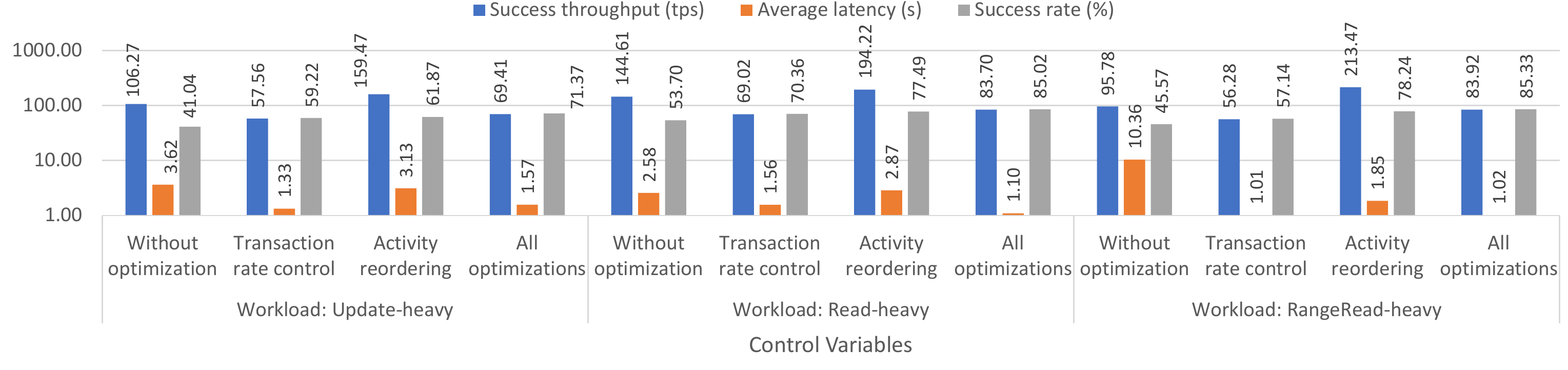}
\captionsetup{justification=centering} 
\captionsetup{labelfont={bf}}
\caption{Synthetic workloads with Fabric++}
\label{fpp}
\end{figure*}

\emph{Electronic Health Records (EHR)}:  
In this use-case, three optimizations were recommended: activity reordering, process model pruning and transaction rate control (Figure~\ref{ehrgraph}). Activity reordering for the read activities resulted in a 60-65\% improvement in throughput and success rate. When the smart contract was updated to prune illogical paths (revoke access to records without granting access), we observe around 43\% increase in throughput and success rate. After applying transaction rate control, a 69\% increase in success rate was observed. All optimizations applied together also improve the performance.

% \begin{figure}[ht]
% \centering
% \includegraphics[width=0.6\columnwidth]{figures/dvgraph.pdf}
% \caption{Digital voting use-case}
% \label{dvgraph}
% \end{figure}
\emph{Digital Voting (DV)}\label{dvresults}: 
In this use-case, two optimizations were recommended: transaction rate control and data model alteration. The results are shown in Figure~\ref{dvgraph}. High failure rates were detected for periods when the \texttt{Vote} transactions were frequent. After applying transaction rate control, a slight improvement of 11\% in throughput was observed. The hotkeys were detected and most frequently used by the \texttt{Vote} function resulting in a recommendation to alter the data model. We analysed the smart contract and observed that \texttt{partyID} was used as the key for the vote function which is invoked by multiple voters during the voting phase. We redesigned the smart contract such that \texttt{voterID} is assigned as the primary key. Since voters are restricted to a single vote, we observe 100\% success rate with this new smart contract because there are no more transaction dependencies. We also observe an improvement in the performance when both optimizations are applied together.

%, the throughput is less than the throughput obtained by applying only data model alteration. This is because transaction rate control decreases the throughput.

 \subsection{Loan Application Process (LAP)}\label{lapresults}
% \begin{figure}[ht]
% \centering
% \includegraphics[width=\columnwidth]{figures/lapgraph.pdf}
% \caption{Loan application process use-case}
% \label{lapgraph}
% \end{figure}

The optimization recommended for the LAP use-case was data model alteration (Figure~\ref{lapgraph}). The \texttt{employeeID} 1 had a high key frequency since this employee processed the highest number of loan applications. We then re-implemented our smart contract and assigned \texttt{applicationID} as the key and modeled the value as a structure that includes \texttt{employeeID}, \texttt{loan amount}, \texttt{loan type} and \texttt{loan status}. This new implementation helped to remove the hot key and yielded more than 50\% improvement in throughput and success rate for both the lower and higher send rates.

\subsection{Fabric Extensions \label{fabricextensions}} 
As a holistic recommendation approach, our work lies orthogonal to existing Fabric optimizations in the literature. In this section, we demonstrate how our approach works on top of two optimized extensions of Fabric: FabricSharp~\cite{10.1145/3318464.3389693} and Fabric++~\cite{Sharma:2019:BLB:3299869.3319883}. Both implement different transaction reordering strategies that mitigate MVCC read conflicts. The Fabric++ scheduler is integrated in the FabricSharp implementation~\cite{fsgit} and we use this for our experiments. We executed the synthetic workloads on both and then used \textsf{BlockOptR} to generate recommendations. The literature says FabricSharp increases endorsement policy failures and is less performant for insert-heavy workloads while Fabric++ is least performant with an update-heavy, read-heavy and range-read-heavy workloads~\cite{10.1145/3448016.3452823}. Therefore, we execute these specific experiments shown in Figures~\ref{fsgen} and ~\ref{fpp} with the synthetic workloads. Activity reordering, transaction rate control and endorser restructuring were recommended and by implementing these recommendations, we observe up to a 55\% increase in throughput and 46\% increase in success rate (Figure~\ref{fpp}: RangeRead-heavy workload). Our experiments with these Fabric extensions show that even with effective system-level optimizations, Fabric can still benefit from optimizations at all levels of abstraction.

%The FabricSharp implementation provides a configuration setting to switch to the Fabric++ reordering strategy which we used to evaluate Fabric++. We executed the same .... \textsf{BlockOptR} generated . Implementing this increased the throughput by 26\% and the success rate by 6\% (Figure~\ref{ff}).   

%\footnote{We setup FastFabric at a smaller scale compared to the literature which has a much higher throughput. We also faced a setup issue that we have submitted to the developers\apostrophe ~ git~\cite{ffissue} which prevented more extensive experiments.}. 

% \begin{figure}[ht]
% \centering
% \includegraphics[width=0.8\columnwidth]{figures/fabricsharp_scm.pdf}
% \caption{SCM with FabricSharp}
% \label{fsscm}
% \end{figure}

% \begin{figure}[ht]
% \centering
% \includegraphics[width=0.8\columnwidth]{figures/fabricsharp_generator.pdf}
% \caption{Generator with FabricSharp}
% \label{fsgen}
% \end{figure}

% \begin{figure}[ht]
% \centering
% \includegraphics[width=0.8\columnwidth]{figures/fastfabric.pdf}
% \caption{FastFabric experiments}
% \label{ff}
% \end{figure}

% \begin{figure}[ht]
% \centering
% \includegraphics[width=0.8\columnwidth]{figures/fastfabriclogscale.pdf}
% \caption{FastFabric experiments: Log scale graph}
% \label{ff}
% \end{figure}

\section{Lessons Learned and Limitations}
\label{sec:discussion}
We demonstrated that \textsf{BlockOptR} is capable of effectively recommending suitable optimization strategies. Further, we also explained how to implement these optimizations and quantified the performance improvements after implementation. This section discusses the insights we gained from our experiments.

\textbf{User level optimizations.} Activity reordering was one of the most frequently recommended optimizations in our experiments. We highlight use-cases such as SCM where such reordering can be applicable. Our model pruning recommendation emphasizes that identifying incompetencies in the process model can lead not only to efficient process execution but also improve the performance of the underlying system. Load shedding or queuing is often employed when systems cannot handle the workload. Using our recommendations, specific activities and time periods can be identified where such rate control techniques are most effective. For example, rate control is recommended for the \texttt{Vote} activity in the digital voting use-case. Therefore, instead of system-wide rate control, only the specific clients that deal with the identified activities need to employ rate control techniques. 

\textbf{Data level optimizations.} These optimizations show how the design of the smart contract and the data model significantly influence the performance. The smart contract is initially designed with a specific process model in mind. However, we understand how the smart contract is being used in practice by analyzing the blockchain logs. \textsf{BlockOptR} pinpoints functions and keys that cause bottlenecks which in turn helps the smart contract developer to make appropriate modifications.

\textbf{System level optimizations.} Setting the endorsement policy is a management decision that often excludes discussions with the technical team designing the blockchain. Our recommendations highlight the need to bring together management and technical discussions to decide optimal configuration settings. Further, we also demonstrate the need to verify whether the policy is being used effectively. For example, even if the policy defines the equal distribution of endorsements, the clients may send their transactions in a skewed manner. In such instances, we recommend enforcing a management measure, such as dividing the endorsers equally among the clients such that clients of one organization only send transactions to specific endorsers. The compliance with such measures can also be checked by \textsf{BlockOptR}. Block size optimization is frequently discussed in the literature and associated with the transaction rate of a system~\cite{10.1145/3448016.3452823, 8526892,istvan2018streamchain}. Instead of system-level changes such as using transaction rate monitors, we derive the transaction rate and the actual block size from the log. This helps to understand traffic patterns over time and find reasonable block size settings. While the literature mainly focuses on optimizing the peers and ordering service components of Fabric~\cite{gorenflo2019fastfabric, 10.1145/3318464.3389693, Sharma:2019:BLB:3299869.3319883}, our client-related recommendations highlight the need to focus on client-side optimizations as well.

\textbf{Technology Independence.}
Our multi-level recommendation approach is demonstrated using the Fabric blockchain. Technology independence is difficult to attain due to the vast implementation variations between the numerous blockchain systems and the corresponding differences in the contents of the distributed ledger. However, we draw attention to specific examples which can guide future researchers to translate our approach to other blockchain systems. In Quorum, the block time or mining frequency has a linearly proportional influence on the transaction latencies~\cite{https://doi.org/10.48550/arxiv.1809.03421} which is analogous to our block size adaptation recommendation strategy. Also, Corda has the concept of notaries to attest transactions where distributing the transactions over multiple notaries is expected to improve the throughput~\cite{cordanotaries}. This is again comparable to our endorsement restructuring recommendation. Further, there are numerous gas-fee reduction and vulnerability detection strategies for Ethereum smart contracts in the literature~\cite{9569819} which translate to our recommendations at the data level. Tools like Lorikeet and Caterpiller automate the conversion and execution of process models as Ethereum smart contracts, which would make it easier to implement the user-level optimizations that we recommend~\cite{https://doi.org/10.48550/arxiv.1808.03517, DBLP:conf/bpm/TranLW18}.    

\textbf{Limitations.}
The optimizations recommended by \textsf{BlockOptR} need to be manually implemented by the user. A self-adaptive system with a feedback loop that automatically implements the recommendations is possible. However, in an enterprise scenario, for many of the optimizations such as endorser restructuring, activity reordering, and process model pruning, management level approvals might be required before implementation. Additionally, for applications that do not follow a specific process model, the event logs can be misleading. In such scenarios, user-level optimizations such as activity reordering and process model pruning are not relevant. Therefore, domain knowledge about the use-case is required for implementing the recommended optimizations appropriately. Further, our implementation of some of the optimizations such as transaction rate control are trivial in such benchmarking scenarios and do not account for real-world overheads. However, the implementations are mainly for demonstrative purposes. Our work focuses on the multi-level recommendation approach used by \textsf{BlockOptR} rather than the implementation of the optimizations. Finally, our experiments without and with the recommended optimizations are done on similar workloads generated with the same input parameters, i.e., we assume a continued trend in the pattern of the workload after the optimizations are applied. However, in scenarios where the workload fluctuates or the optimization implementation is delayed, \textsf{BlockOptR} may need to be re-executed to generate new recommendations. 

%Finally, our experiments without and with the recommended optimizations are done on similar workloads generated with the same input parameters, i.e., we assume a continued trend in the pattern of the workload after the optimizations are applied. However, in scenarios where the workload varies, \textsf{BlockOptR} may need to be re-executed to generate new recommendations.

\section{Related Work}
\label{sec:related}
%Blockchain optimizations—ours orthogonal
The literature proposes various Fabric optimization strategies such as transaction reordering~\cite{Sharma:2019:BLB:3299869.3319883, 10.1145/3318464.3389693, 10.1145/3448016.3452823}, block size optimizations~\cite{istvan2018streamchain, 10.1145/3448016.3452823}, CRDTs~\cite{10.1145/3361525.3361540}, and parallelizing various components~\cite{gorenflo2019fastfabric}. Our work lies orthogonal to such optimization strategies and focuses on an optimization recommendation approach. We demonstrate how our recommendations can be used along with two of the literature's optimization strategies to improve performance further.

%DB tuning index recommendations
There is also extensive research in the database community on index and query optimizations that include self-tuning systems as well as recommendation systems~\cite{839397, 7495648, 10.1007/978-3-642-02279-1_2, 10.1007/978-3-319-93803-5_1, 10.5555/1325851.1325856}. Though we can draw parallels from these research, our work focuses on blockchain-specific optimization recommendations. Different configuration settings (such as block size and endorsement policy) and the concept of smart contracts introduce new dimensions to the recommendation approach, which are not required for databases.

%Process level – Process mining papers, DB,
There is ongoing research on applying process mining techniques on blockchains to derive process-level insights~\cite{klinkmuller2019mining, muhlberger2019extracting, duchmann2019validation, hobeck2021process}. Klinkm{\"u}ller et al.~\cite{klinkmuller2019mining} and M{\"u}hlberger et al.~\cite{muhlberger2019extracting} describe different approaches to extract process data from the Ethereum blockchain. Hobeck et al.~\cite{hobeck2021process} use process mining on an Ethereum-based betting application to identify shortcomings in the application. Process mining on blockchains currently only focuses on permissionless blockchains as they are publicly accessible. However, deriving and studying the process model is equally critical for private blockchains, and therefore, our work contributes to this less explored area of research. Further, unlike the related work, we focus on using process mining for recommending blockchain optimization strategies. We only found a single paper that uses permissioned blockchains, where Duchmann et al.~\cite{duchmann2019validation} extract process data from Fabric and detect semantic errors in a smart contract. Though our work is comparable, we extract not only the process data but also blockchain-specific attributes from Fabric, derive multiple metrics, and recommend optimization strategies.

There is extensive research in the database community in the domain of data-aware business processes that encourage a business process perspective to database management systems~\cite{ 10.1145/2463664.2467796, 10.1145/1989284.1989286, 10.1007/978-3-540-88873-4_17}. Calvanese et al.~\cite{10.1145/2463664.2467796} comprehensively survey the contributions in this realm and catalog contributions from various fields, including database theory and process management. These works were an important motivation for us to view blockchains from a business process perspective. However, our work brings new contributions since blockchains deal with several other elements apart from data, such as smart contracts and endorsement policies.

\section{Conclusions}
\label{sec:conclusion}
This paper showcases the necessity and effectiveness of having a holistic perspective on blockchain optimizations. We define a multi-level recommendation approach based on several metrics and attributes derived from the blockchain log. We define a total of nine optimizations at the system, data, and user-level of a blockchain. We implement an automated optimization recommendation tool, \textsf{BlockOptR}, based on these concepts. Further, we demonstrate how such optimizations can be implemented to improve the system performance. After implementing the recommended optimizations, we observe an average of 20\% improvement in the success rate and an average of 40\% improvement in latency. We extensively evaluate the system with a wide range of workloads covering multiple real-world scenarios. We hope to inspire enterprises to use our contributions to detect blockchain optimization strategies and to contribute their live blockchain (anonymized) logs for further research in this domain. The \textsf{BlockOptR} tool, all the smart contracts, the workload generation scripts, and all the event logs are available as open-source~\cite{blockprom}. We also plan to extend our tool to include more optimization recommendations.

In terms of future work, we are currently developing a ProM plugin which would provide a user-friendly interface for BlockOptR. Presently, the threshold settings of BlockOptR depend on the business network setup. For example, the rate threshold for our setup was 300 TPS as higher rates led to instabilities, but this can vary for other deployments. Therefore, tuning these thresholds automatically in BlockOptR could be a future extension. Another interesting extension is to define additional attributes that applications can log, thereby providing more data for optimization recommendations. Further, investigating the effect of workload fluctuations and delay in applying the recommendations is another challenging future direction.

%\section{Acknowledgments}
\begin{acks}
This work is funded in part by the Deutsche Forschungsgemeinschaft (DFG, German Research Foundation) - 392214008, and by the Bavarian Cooperative Research Program of the Free State of Bavaria - DIK-2002-0013//DIK0114/02.
%This work is funded by the \grantsponsor{testing} - \grantnum{392214008}
\end{acks}

%\balance
\newpage
%\nolinenumbers

\bibliographystyle{ACM-Reference-Format}
\bibliography{references}


%\bibliographystyle{abbrv}

% \newpage
% \newpage
% \input{content/AllExperiments}

\end{document}